\documentstyle[aas2pp4,flushrt,psfig2]{article}

\def\simless{\mathbin{\lower 3pt\hbox
   {$\rlap{\raise 5pt\hbox{$\char'074$}}\mathchar"7218$}}} 
\def\simgreat{\mathbin{\lower 3pt\hbox
   {$\rlap{\raise 5pt\hbox{$\char'076$}}\mathchar"7218$}}}

\begin{document}

%
%

\title{RADIO SCINTILLATION DUE TO DISCONTINUITIES IN THE 
INTERSTELLAR PLASMA DENSITY}

\author{\sc H.\ C.\ Lambert\altaffilmark{1}}
\affil{Max-Planck-Institut f\"{u}r Radioastronomie \\
Auf dem H\"{u}gel 69, D-53121 Bonn, Germany}
\author{\sc B.\ J.\ Rickett\altaffilmark{2}}
\affil{Department of Electrical \& Computer Engineering \\
University of California San Diego, La Jolla, CA 92093-0407}
\author{\sc DRAFT: \today}

\altaffiltext{1}{e-mail: hlambert@mpifr-bonn.mpg.de}
\altaffiltext{2}{e-mail: rickett@ece.ucsd.edu}

%
%

\begin{abstract}
We develop the theory of interstellar scintillation 
as caused by an irregular plasma having a power-law 
spatial density spectrum with a spectral exponent of 
$\beta = 4$ corresponding to a medium with abrupt 
changes in its density.  An ``outer scale'' is included 
in the model representing the typical scale over which the 
density of the medium remains uniform.  Such
a spectrum could be used to model plasma shock 
fronts in supernova remnants or other plasma discontinuities.  
We investigate and develop
equations for the decorrelation bandwidth
of diffractive scintillations and the refractive 
scintillation index and compare our results with
pulsar measurements.  We consider both a medium concentrated
in a thin layer and an extended irregular medium. 
We conclude that the $\beta = 4$ model gives satisfactory 
agreement for many diffractive measurements,
in particular the VLBI meaurements of the structure function 
exponent between 5/3 and 2.  However,
it gives less satisfactory agreement for the 
refractive  scintillation index than does the 
Kolmogorov turbulence spectrum. The comparison suggests that 
the medium consists of a pervasive background distribution of
turbulence embedded with randomly placed 
discrete plasma structures such as shocks or HII regions.
This can be modeled by a composite spectrum following
the Kolmogorov form at high wavenumbers and steepening 
at lower wavenumbers corresponding to the typical 
(inverse) size of the discrete structures. 
Such a model can also explain the extreme scattering events.
However, lines of sight through the enhanced scattering
prevalent at low galactic latitudes are accurately described by
the Kolmogorov spectrum in an extended medium and do not appear
to have a similar low-wavenumber steepening.
\end{abstract}


%
%

\section{INTRODUCTION}

It is well known that radio waves propagating in the 
interstellar medium (ISM) are scattered by the irregularities 
in the Galactic electron density.  The scattering in 
turn gives rise to a number of observable phenomena.  
Among others, these include angular broadening
and intensity fluctuations (both in time and frequency) of
compact radio sources.   
While a nuisance in many radio astronomical observations, these 
phenomena can be used to investigate the nature of the 
irregularities in the interstellar plasma density.  These density
irregularities are in turn believed to follow the fluctuations
in the interstellar kinetic and magnetic energies.  
Ideally one would like to invert observations of radio scintillation
and scattering to determine the statistics of the plasma density.
As noted by Narayan (1992) this inverse problem is not well
posed and one must rely on modeling methods.
A complete prediction of scintillation observables requires
an {\em a priori} knowledge of both the form of the spatial power 
spectrum of the electron density fluctuations and 
the distribution of the scattering material along the line of 
sight.  For a given profile of the distribution of the scattering
material, one may compare observations and predictions and 
so constrain the functional form of the spectrum.  
The power spectrum provides useful insight into the 
physics of the plasma irregularities.
Hence a knowledge of the form of the density
spectrum becomes central in both predicting scintillation
phenomena and understanding the physics of the interstellar plasma.
In this paper, we revisit the investigation of the form of the
density spectrum.  

A commonly used model for the density spectrum has been
based on a power-law model with a large range between
``inner'' and ``outer'' scales (e.g.\ Rickett 1977):
\begin{equation}
P_{N_e}(q) = 
\frac{C_{N_e}^2(z)}{(q^2 + \kappa_o^2)^\beta}
\exp\left[ -\frac{q^2}{4\kappa_i^2} \right]
\, \mbox{.}
\label{eq:extpowerlaw}
\end{equation}
Here $q$ is the magnitude of the three-dimensional wavenumber $\vec{q}$.
  
$C_{N_e}^2(z)$ denotes the strength of fluctuations (with
a weak dependence on distance $z$).  $\beta$ is the spectral
exponent, and $\kappa_i^{-1} = L_i$ and $\kappa_o^{-1} = L_o$
are the inner and outer scales respectively.  For 
$\kappa_o << q << \kappa_i$, we obtain the {\em simple} power-law
model: $P_{N_e}(q) = C_{N_e}^2 q^{-\beta}$.  
Armstrong, Rickett, \& Spangler (1995) have constructed an 
empirical density
spectrum by combining radio scintillation observations in the local
ISM ($\simless$ 1~kpc) with measurements of the differential
Faraday rotation angles and large-scale electron density
gradients.  They have shown the power spectrum to be consistent
with a simple Kolmogorov power-law model ($\beta=11/3$)
over an astronomical 10 orders of magnitude in wavenumber scale 
($10^{-18}$m$^{-1} < q < 10^{-8}$m$^{-1}$). 
The Kolmogorov spectrum in density suggests a turbulent cascade in the
magnetic and kinetic energies. This has lead to several theoretical
investigations of the generation and maintenance of hydromagnetic 
turbulence in the ISM (e.g.\ Pouquet, 1978; Higdon 1984 \& 1986; Biskamp, 1993:
Sridhar \& Goldreich 1994; Goldreich
\& Sridhar 1995 and 1997).  
However, the Armstrong et al.\ study combined observations from many
lines of sight and the scatter among them leaves a substantial
uncertainty in the exponent $\beta$. A list of symbols is given
in Table 3.

In spite of the positive
evidence for the simple Kolmogorov spectrum, substantial
observational inconsistencies remain.  For instance, long-term
refractive intensity scintillations of some pulsars
have modulation indices as much as a factor of 2 larger
than predicted by the simple Kolmogorov model (cf.\
Gupta, Rickett, \& Coles 1993).  Other discrepancies are revealed
in the diffractive dynamic spectra of pulsars.  On some 
occasions, periodic fringes are observed, which are not 
predicted by the simple Kolmogorov model (cf.\ Roberts \&
Ables 1982; Cordes \& Wolszczan 1986; Rickett, Lyne, \& Gupta 1997;
Gupta, Bhat \& Rao, 1999); in addition, sloping bands 
in the dynamic spectra often persist
longer than predicted by the model (cf.\
Gupta, Rickett, \& Lyne 1994; Bhat, Rao \& Gupta, 1999b
Bhat, Gupta \& Rao, 1999c).  
Further for some pulsars, the decorrelation bandwidth has 
larger amplitude variations than predicted for the Kolmogorov 
spectrum (Bhat, Gupta \& Rao, 1999c). Such anomalies
suggest the presence of large refractive structures giving rise
to the focusing and defocusing of the scattered ray bundles.  The
interference of the ray bundles
can also explain the occasional fringes observed 
in the dynamic spectra of some pulsars.
Given the relatively frequent occurrence of such events, one can 
ask whether they can be considered as mere occasional anomalies
or should be considered as a widespread phenomenon intrinsic to the
spectrum on a grand scale.  

The observational inconsistencies 
suggest the need for an enhancement in the power on the
large ``refractive'' (10$^{11}$~m to 10$^{12}$~m)
spatial scales relative to the power on the small ``diffractive'' 
(10$^7$~m to 10$^8$~m) scales.  
There are several means by which this ratio may be enhanced.  
One is to include the inner scale cut-off in the
density spectrum, which reduces the power at small scales
(Coles et al.\ 1987); 
these authors  proposed inner scales of $10^8-10^9$ m., 
though this does not correspond to any obvious physical scale.
Physically, the inner scale corresponds to the scale at
which the turbulent cascade dissipates and becomes a source
of heating for the plasma (Spangler, 1991).  
The value of the inner scale is largely unknown. Using different 
methods, Spangler \& Gwinn (1990), 
Kaspi \& Stinebring (1992), and Gupta et al.\ (1993) have reported 
values for the inner scale ranging from $10^4$ to $10^9$ meters. 
In proposing the smaller values, Spangler \& Gwinn (1990) argued
that the inner scale is the larger of the ion inertial length,
$L_i \equiv V_A/\Omega_i$ (where $V_A$ is the Alfv\'{e}n speed, 
and $\Omega_i$ is the ion cyclotron frequency), 
and the ion Larmor radius, $r_i \equiv v_{th}/\Omega_i$ 
(where $v_{th}$ is the ion thermal speed);
they obtained  parameters for the warm ionized medium in reasonable
agreement with observations. In a recent discussion, Minter \&
Spangler (1997) have suggested ion-neutral collisional damping and
wave-packet steepening as possible dissipation mechanisms for
the turbulence in the diffuse ionized gas, which 
would make the mean-free path for ion-neutral collisions 
a possible value for the inner scale. However, this is thought 
to be larger
than the maximum values proposed to explain the observations,
making it a less convincing dissipation mechanism.
Observationally, the Kolmogorov model with a large
inner scale predicts refractive modulation indices consistent with pulsar 
measurements (Gupta et al.\ 1993).   It has also been proposed
to explain the occasional periodic fringes, 
with an inner scale on the order of the Fresnel scale 
(Cordes, Pidwerbetsky, \& Lovelace 1986; Goodman et al.\ 1987).  
However, Rickett et al.\ (1997) reported a fringe event for 
the pulsar B0834+06 that could not be explained as
the effect of a large inner-scale spectrum.  The event 
requires similar conditions to those needed to explain
the extreme scattering events (Fiedler et al.\ 1987).

Another way to enhance the ratio of the power between the 
refractive and diffractive scales in the spectrum is to steepen the 
spectrum---with spectral exponents $\beta > 4$ (Blandford \& Narayan
1985; Goodman \& Narayan 1985; Romani, Narayan, \& Blandford 1986).  
While power-law spectra with $\beta \sim$ 11/3 have a turbulence 
connotation, spectra with $4<\beta<6$ might involve
some forms of turbulence, but are also consistent with a 
distribution of non-turbulent structures with a range of 
spatial scales.  
Such steep spectra with $4 < \beta < 6$ predict 
refractive modulation indices
close to unity (Goodman \& Narayan 1985), which is substantially larger than 
the range 30 to 40 \% observed from the nearby pulsars.  
On this basis Rickett \& Lyne (1990) and 
Armstrong et al.\ (1995) have rejected spectra steeper than 4 for the
interstellar plasma.  However, the special case of $\beta=4$ has been
given little attention.  We can conceive of a  power-law model 
with spectral exponent $\beta=4$ given by:
\begin{equation}
P_{N_e}(q;z) = \frac{C_{N_e}^2} {\left(q^2+\kappa_o^2\right)^2}
\, \mbox{.}
\label{eq:beta4}
\end{equation}
Hereafter, we refer to this model as the ``$\beta=4$ model.''
Blandford \& Narayan (1985) briefly discussed this special case
without including a cut-off at low wavenumbers. 
It is interesting to note that even though the Kolmogorov spectral 
exponent $\beta=11/3$ is very close to 4, the $\beta=4$ model 
has a very different physical implication.  
Its physical origin has been rarely discussed, and it has
not been formally compared with observations.  Physically, this 
spectrum suggests the random distribution in location and 
orientation of discrete discontinuous objects across the line 
of sight.  An ``outer'' scale, $L_o=\kappa_o^{-1}$,
is included to account for the typical size of such
objects.  The ``inner'' scale here would correspond to the 
scale of the sharpness of a typical discontinuity.  We assume the
``inner'' scale to be smaller than the diffractive scale 
of scintillations, and hence has no significant effect on the
scintillations.  The $\beta=4$ model could
characterize stellar wind boundaries, supernova shock fronts, 
sharp boundaries of HII 
regions at the Str{\"o}mgren radius, or any plasma ``cloud'' with 
sufficiently sharp boundaries (transition regions shorter than the 
diffractive scintillation scale) which may cross the line of sight. 
Note that, though turbulence is not necessarily implied by the model,
strong turbulence which has steepened to form shocks would also
be described by the  $\beta=4$ model.
In these models a single discrete object crossing
the line of sight could also explain the ``extreme 
scattering events'' observed in the flux density variations of 
extra-galactic sources (e.g.\ Fiedler et al.\ 1987).
We note that the analysis of such events have been primarily based 
on geometrical optics involving a single ``cloud'' (Goodman et 
al. 1987; Cordes et al.\ 1986; 
Cordes \& Wolszczan 1986; Roberts 
\& Ables 1982; Ewing et al.\ 1970).  Our analysis of 
the $\beta=4$ model includes the ``wave optics'' effects when the 
line of sight passes through very many such clouds.  The ISM
is assumed to consist of a random assembly of discrete
structures with abrupt density steps which may be independent
of each other.  If the $\beta=4$ model were compatible with all
of the scintillation observations, it would remove the
implication of interstellar plasma turbulence with
an inertial range spanning as much as 10 orders of
magnitude in scale, which has become the canonical
model for ISS phenomena. It would, however, still be consistent 
with turbulence that has steepened into shocks.

We start with a simple derivation of the 
$\beta=4$ model in section \ref{sec:deriv}.  
In section \ref{sec:structfn},
we give equations for the wave structure function
for the $\beta=4$ model and compare predictions
In section \ref{sec:decbw},  the variation of the diffractive
decorrelation bandwidth with frequency is used to test the
$\beta=4$ model against the simple Kolmogorov model and the
Kolmogorov model with an inner-scale (hereafter, the ``inner-scale''
model).
In section \ref{sec:mr}, the observed variation of the refractive
scintillation index with the normalized diffractive decorrelation
bandwidth is compared with theoretical predictions
for the simple Kolmogorov, inner-scale, and $\beta=4$ models. 
We use theoretical results from a previous 
paper (Lambert \& Rickett 1999; hereafter, LR99), 
in which we developed the theory of diffractive scintillations
in a medium modeled by these spectra, and we add
details of the theory for refractive scintillation in Appendix A. 
In Appendix B, we discuss the use of the
square of the second moment to approximate the intensity correlation 
function for diffractive scintillation with $\beta=4$.
In section \ref{sec:conclusion} we give a discussion and our 
conclusions.

%
%

\section{THE SPECTRUM FOR RANDOM \newline DISCONTINUITIES IN DENSITY}
\label{sec:deriv}

In this section, we give a simple derivation of the $\beta=4$ model.  
Consider first a random distribution in space 
of identical plasma structures (blobs). 
The resulting electron density may be described as the 
convolution of a three-dimensional Poisson point
process with the density profile of one individual blob.  
This is the spatial analog of shot noise, in which each charge carrier 
contributes the same temporal profile of current arriving
at randomly distributed times.  
In the wavenumber domain, the power spectrum of the electron density
is then given by the spectrum of the Poisson point process 
multiplied by the squared magnitude of the
Fourier transform of one blob.   The spectrum of the
Poisson point process is a constant equal to
the number density, $n_0$, of the blobs in space (Papoulis 1991). 
Thus the power spectrum of the medium
follows the same shape as that of an individual blob.
A similar description applies to water droplets in a fog
and many naturally occurring media, as discussed by Ratcliffe (1956).
If the blobs are asymmetrical and randomly oriented, one must also average
the spectrum over the possible angles of orientation. 

The basic feature of
structures with discontinuous boundaries is that
their power spectrum has an asymptotic behavior at large wavenumbers
that varies as (wavenumber)$^{-4}$. 
This is the spatial analog of the idea that the temporal spectrum
of any pulse shape with an abrupt rise or fall has a high-frequency
asymptote as (frequency)$^{-2}$.  The prime example
is a rectangular pulse, for which the spectrum is a sinc-squared function, 
whose high-frequency envelope follows this law. 
The simplest spatial example is spherical blobs of radius
$a$ with uniform density $f_0$ inside and zero outside;
clearly, an added uniform background density does not
affect the power spectrum.  The three-dimensional 
spatial Fourier transform of an \it isotropic \rm
function $f(r)$ of radial distance $r$ is given by
(Tatarskii 1961):
\begin{equation}
F(q) = \frac{1}{2\pi^2q}
\int_0^{\infty} \, f(r) \sin(q r) r \, dr
\, \mbox{,}
\label{eq:isofourier}
\end{equation}
where $q$ is the magnitude of the wavenumber.
For any function $f(r)$ that falls to zero beyond a radius $a$,
the integral will vary as $1/q$ in the high-wavenumber limit
$q a \gg 1$, making $F(q) \propto 1/q^2$.
For the spherical blobs of size $a$ we have:
\begin{equation}
F(q) = \frac{f_0}{2\pi^2 q}
\left\{
- \frac{a \cos(q a)}{q} +
  \frac{\sin(q a)}{q^2}
\right\}
\, \mbox{.}
\label{eq:spherefourier}
\end{equation}
For the squared magnitude of the Fourier transform,
for very small and large values of $q a$:
\begin{equation}
| F(q) |^2 =
\left\{
\begin{array}{lr}
{\displaystyle
f_0^2 a^6 /(36 \pi^4) \, \mbox{,} 
} &
q a << 1 \, \mbox{,}
\\ [0.25in]
{\displaystyle
 q^{-4} \; a^2 f_0^2 /(8\pi^4) \, \mbox{,} 
} &
q a >> 1 \, \mbox{.}
\\
\end{array}
\right.
\label{eq:magsqspherefourier}
\end{equation}
The spectrum of the electron density is then
$n_0 |F(q)|^2$.  We can approximate this by equation (\ref{eq:beta4}),
which exactly matches equation (\ref{eq:magsqspherefourier}) at small
and large values of $qa$, 
with  $C_{N_e}^2 = n_0 a^2 f_0^2 / (8 \pi^4)$ and
$\kappa_o^{-1} = L_o = (2/9)^{0.25} a$.

There are evident generalizations to make the model more 
like a real medium. The spheres could have a probability
distribution for their radii, which would weight the
average of the square of equation (\ref{eq:spherefourier}).
The spheres could have smoothly varying density inside and an abrupt 
boundary at radius $a$.  As with the pulse example, the
high-wavenumber behavior would not change. 
More realistic models would include  
anisotropic structures with random orientations and
a distribution of scales.  For example, consider a 
circular cylinder as a model for an anisotropic blob, with a
particular radius and thickness.  Its Fourier transform can be 
computed with respect to its major axes.  
This can then be transformed to tilted axes and the result averaged over
an isotropic distribution of angles of tilt.
Evidently the result will also be isotropic; we have solved this
approximately and find a spectrum with the same high-wavenumber 
\it asymptotic \rm form as equation
(\ref{eq:beta4}), where the ``outer scale,'' $L_o$,
is the smaller of the thickness and radius.  
This result could be averaged over a probability distribution of
radii and thicknesses and a similar conclusion reached, with $L_o$
as a weighted average of the smallest dimension.  
Planar sheets at random angles are modeled as very flat cylinders
and provide a crude representation of a distribution of shocks. 
Another simplified model would be ellipsoids 
of different sizes, eccentricities, and orientations, and
again we expect a similar result. 

We conclude that the $\beta=4$ spectrum is an approximate description
of the power spectrum for a medium with a random
distribution of discontinuous structures, in the limit for wavenumbers
greater than the reciprocal of the 
smallest dimension across the structures.
We note that for wavenumbers at or below about $\kappa_o = L_o^{-1}$ 
the functional form depends on the detailed shapes, 
though it will be much less steep than $\kappa^{-4}$. Thus
equation (\ref{eq:beta4}) will not be generally 
applicable at the lowest wavenumbers.
Similarly, in practice there is a finite scale over which 
the density jump occurs (e.g.\ shock thickness).  This can be modeled
by convolving the idealized discontinuous structures with a 
suitable narrow, say, Gaussian function which thus provides a
very high-wavenumber cut-off beyond which the spectrum falls
rapidly to zero.  Consequently, we consider a spectrum with an
inverse fourth power law between inner and outer scales,
though these do not have 
the same connotations as
inner and outer scales in a turbulent cascade.
We do not discuss the physical origin of the supposed 
discontinuous plasma structures beyond the ideas
formulated in the Introduction,
such as supernova shock fronts, 
stellar wind boundaries or sharp boundaries of HII 
regions at the Str{\"o}mgren radius.  There needs to be
no physical coupling between the individual blob structures, 
and there is no turbulent cascade implied by the $\beta=4$ spectrum.
However, we note that shocks developing from
the steepening of very strong turbulence could also 
be described by the model, and so if successful, the
model would not rule out such strong turbulence,
though in this case the $\beta=4$ range does not correspond 
to an inertial cascade as does the 11/3 spectrum.

In the following sections, we give the necessary 
scintillation theory needed to develop
equations for the scintillation observables corresponding
to the $\beta=4$ model, which we then compare with
ISS observations. We concentrate on measurements which
are sensitive to the form of the density spectrum and are 
relatively insensitive to the distribution of scattering material
along the line of sight.

%
%

\section{PHASE STRUCTURE FUNCTION}
\label{sec:structfn}

%
%
%

Central to the description of the second moment and
the coherence function for intensity is the phase structure 
function.  The longitudinal gradient of the phase structure function
is related to the electron density spectrum through (cf.\ Coles et al.\ 
1987):
\begin{eqnarray}
D_{\phi}'(s,z;\nu) & = & \frac{8\pi^2 r_{\rm e}^2 c^2}{\nu^2}
\int_0^{\infty} P_{N_e}(\kappa, q_z=0;z)  \nonumber \\
& & \times \left[1-J_0(\kappa s)\right] \kappa d\kappa
\, \mbox{.}
\label{eq:long_grad_phase_strucfunc}
\end{eqnarray}
Here $c$ is the speed of light, $r_{\rm e}$ is the 
classical electron radius, and $\nu$ is the radio
frequency. For a plane wave incident on a scattering
medium, the line of sight integral of
$D_{\phi}'(s,z;\nu)$ gives the structure function
of the geometric optics phase, also called the 
{\em wave} structure function, $D_{\phi}(s)$.
For the spherical wave geometry applicable to
a pulsar at distance $L$, the separation $s$ at the observer's 
plane is projected back to the point of integration,
giving $D_{\rm S}(s) = \int_0^L D_{\phi}'(sz/L,z;\nu) \, dz$.
The integrals are evaluated explicitly
in equations (13), (17), and (21) of LR99
for a uniform scattering medium with
the simple power-law, inner-scale, and $\beta=4$ 
spectral models, respectively.
In the extended scattering geometry, 
the field coherence length at the observer, $s_0$, 
is defined by: $D_{\rm S}(s_0)=1$.
For a screen of thickness $\Delta z$ at distance $z_p$ 
from the pulsar, the wave structure function at the observer
becomes $D_{\rm S}(s) =  D_{\phi}'(sz_p/L,z_p;\nu) \Delta z$,
which has an explicit dependence on the screen location.
In LR99, we found it useful to define a coherence scale 
$s_{\rm 0,scr}$ for the screen by its phase structure function 
$D_{\phi}(s_{\rm 0,scr})= D_{\phi}'(s_{\rm 0,scr},z_p;\nu) \Delta z = 1$,
which is independent of the screen location.
Note that in applying equation (\ref{eq:long_grad_phase_strucfunc}),
we are assuming that the screen thickness, $\Delta z$, is much greater
than the outer scale $L_o$ and small compared to total distance $L$.

Here, we repeat the equation for the phase structure function
of the $\beta=4$ model in a thin layer (screen).  
Substituting the $\beta=4$ model for 
the density spectrum into equation (\ref{eq:long_grad_phase_strucfunc}),
integrating, and multiplying by the screen thickness,
$\Delta z$, we obtain:
\begin{equation}
D_{\phi}(s) = \frac{4 \pi^2 r_{\rm e}^2 c^2 SM}{\kappa_o^2 \nu^2}
\left[1-(\kappa_o s)K_1(\kappa_o s)\right]
\, \mbox{,}
\label{eq:struc4exact}
\end{equation}
where $K_1$ is the first order modified Bessel function of the 
second kind, and $SM = C_{N_e}^2 \Delta z$
is the scattering measure.  
Equation (\ref{eq:struc4exact}) may be 
approximated by the following logarithmic expression:
\begin{equation}
D_{\phi}(s) = \frac{\pi^2 r_{\rm e}^2 c^2 SM}{\nu^2}
s^2 \ln\left[1+\frac{4}{(\kappa_o s)^2}\right]
\, \mbox{,}
\label{eq:struc4}
\end{equation}
which matches the full Bessel expression at
both large and small values of the argument.
The detailed shape, near where the structure function flattens,
is governed by the shape of the low-frequency turnover
in our model spectrum, equation (\ref{eq:beta4});
however, as noted above, the turnover shape depends on details of the 
density profiles in the plasma blobs which are not constrained
in our density model.

The logarithmic structure function for the $\beta=4$ model
has far-reaching consequences.  As the outer scale
becomes larger, the structure function approaches
a square-law.  However, due to the presence of the logarithm,
this occurs only slowly; that is, only 
at the limit as $s$ goes to zero does the
$\beta=4$ model structure function become exactly square-law.
A square-law structure function for the medium 
is a convenient mathematical model, but as noted in LR99, in a real
medium the structure function must eventually saturate 
at very large separations.  Nevertheless, the square-law 
structure function has been widely
used since it provides a valid approximation for scales smaller than the 
reciprocal of the wavenumber at which the spectrum is cut-off.
The fact that the $\beta=4$ model
structure function deviates from a pure square-law, even for
very large values of the outer scale, makes this model interesting
and an independent investigation worthwhile. 

%
\begin{figure}[htbp]
\psfig{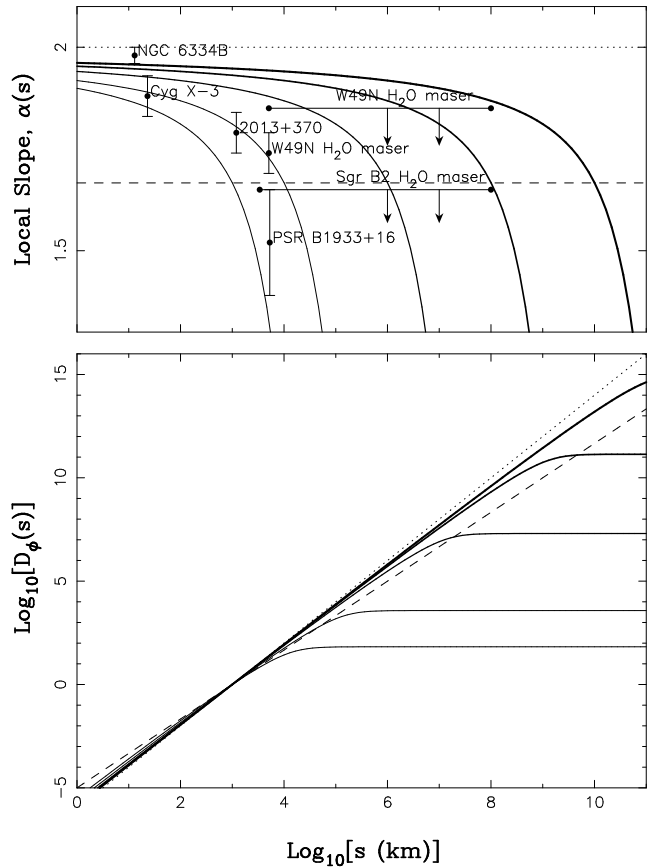}
\caption{
Local logarithmic slope, $\alpha(s)$, of the phase structure function 
(upper panel) and the phase structure function (lower panel) for the 
$\beta=4$ model.  We have set $s_0=1000$ km for all 
curves, and the solid lines correspond to different values
of the outer scale, $L_o$.  In going from the thinnest to thickest
solid lines, the outer scales are: $10^4$, $10^5$, $10^7$, $10^9$,
and $10^{11}$ km, respectively. 
The thin dashed line corresponds to the simple Kolmogorov 
model with logarithmic slope of 5/3, and the thin dotted line 
corresponds to the square-law structure function.  The points in the 
upper panel correspond to measured local logarithmic slopes of the 
phase structure function obtained from VLBI
(Spangler \& Gwinn 1990).}
\label{fig:slope}
\end{figure}

In Figure \ref{fig:slope}, we have plotted the structure function
of the $\beta=4$ model and its effective local logarithmic slope,
$\alpha(s)$, versus the transverse spatial lag, $s$, 
for various values of the outer scale, $L_o$
(see equation 20 of LR99). To illustrate the shapes, 
we have chosen $s_{\rm 0,scr}=1000$~km for all curves.  
The thin dashed lines correspond to the simple Kolmogorov 
model with logarithmic slope of 5/3, and the thin dotted lines 
correspond to the square-law structure function. 
The plot shows how slowly the local slope approaches 2;
even for $s/L_o \sim 10^{-8}$, the slope is 1.95--distinctly
less than 2.
 
The points in the upper panel correspond to measured 
local exponent of the 
wave structure function obtained from VLBI observations, as
tabulated by Spangler \& Gwinn (1990). 
In principle, this method of comparison provides a good test
for the form of the density spectrum.  
These authors presented a similar plot corresponding to 
the inner-scale model, for which $\alpha(s)$ is close to 2.0
for $s$ less than the inner scale and close to 5/3 for
$s$ greater than the inner scale, with a transition
occurring over about a decade in $s$.
They derived estimates 
of the inner scale ranging from 50,000 to 200,000~meters,
for the highly scattered sources that they studied. 
Whereas there is some uncertainty in some of the estimates 
of $\alpha$, there are several examples in which values 
greater than 5/3 are reliably observed (e.g.\ Trotter et al., 1998).  
As can be seen, there is also reasonable agreement with 
the $\beta=4$ model. Assuming the model to be correct,
a separate model-fitting to each observation yields an outer
scale in the range $L_o \sim 3 \times 10^7$ to $10^8$~meters.
These values are much smaller than the parsec scales of supernova remnants and
interstellar clouds.  However based on this 
comparison alone, we are unable to discriminate between
the inner-scale and $\beta=4$ models since they both provide
equally satisfactory agreement with the observations. Hence further
independent tests are needed, and in the following sections, 
we present comparisons with two other scintillation observables
which in the end argue against the $\beta=4$ model as a universal description of 
the ionized ISM.

%
%
\section{DIFFRACTIVE DECORRELATION \newline BANDWIDTH}
\label{sec:decbw}

The diffractive decorrelation bandwidth, $\Delta\nu_d$,
is perhaps the easiest ISS observable to measure. It
is the frequency-difference for a decorrelation to 50\%
of the correlation function for diffractive intensity 
scintillations.  These are usually recorded as a dynamic 
spectrum centered on a given radio frequency, $\nu$, for a 
pulsar. Observations of $\Delta\nu_d$ for one pulsar at a 
wide range of radio frequencies have provided an important 
test of power-law models for the interstellar density 
spectrum on that particular line of sight.  When the 
diffractive scale is far from the inner and outer scales,
one expects $\Delta\nu_d \propto \nu^{2\beta/(\beta-2)}$,
which for the simple Kolmogorov model is $\nu^{4.4}$, in both 
screen and extended medium geometries.  In this section we re-examine the 
published measurements and compare them with theory for the three different 
spectral models. Since the scaling laws are all so steep,
in Figures \ref{fig:vela} and \ref{fig:other}, 
we have plotted the observations logarithmically
as $\Delta\nu_d/\nu^4$.
Overplotted are lines giving the theoretical predictions for our 
three spectral models, with the simple Kolmogorov model as a line of 
slope 0.4.   Before giving our conclusions, we must discuss the 
error bars and various corrections to the data and also the method 
for deriving the theoretical predictions.  

\subsection*{Observations}

%
\begin{figure}[htbp]
\psfig{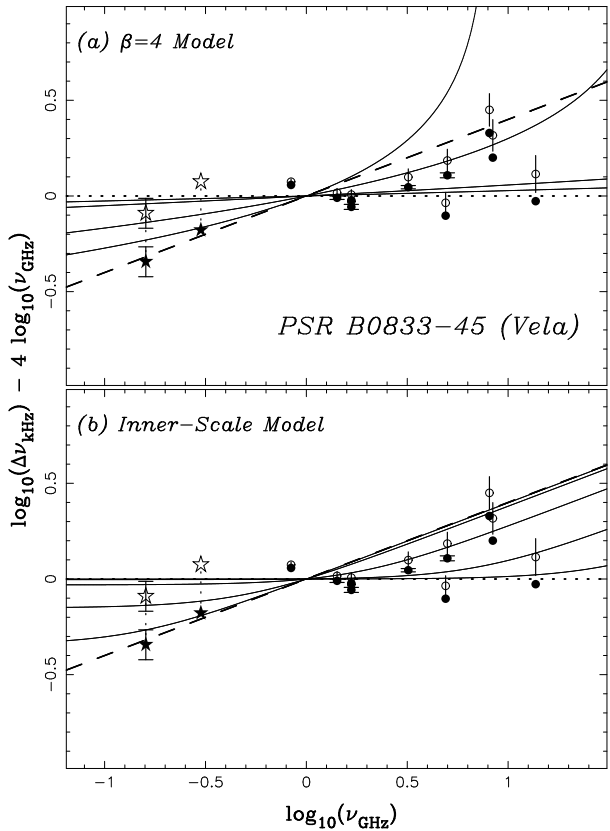}
\caption{
Diffractive Decorrelation bandwidth versus frequency for 
PSR B0833$-$45 (Vela).  For visual aid, a line of slope 4 has 
been subtracted from the log/log plots so that the frequency 
scaling for the square-law structure function model, 
indicated by the thick dotted line, would coincide with 
the abscissa. The filled stars correspond to decorrelation 
bandwidths computed from scattering broadening times using the 
Kolmogorov value for the constant $C_1$, whereas the open stars 
correspond to the $C_1$ from square-law structure function model.
All lines are computed for the extended medium, 
but would be nearly the same for a screen geometry.
The thick dashed line corresponds to the simple Kolmogorov model.  
For the $\beta=4$ model (a), 4 solid lines have been drawn 
corresponding to parameter $\kappa_o s_{\rm 0,1GHz}$ equal to 
$10^{-16}$ (line closest to the square-law structure function model), 
$10^{-8}$, $10^{-3}$, and $10^{-1}$ (line flaring furthest from the 
simple Kolmogorov curve).
For the inner-scale model (b), 5 solid lines have been 
drawn for parameter $\kappa_i s_{\rm 0,1GHz}$, ranging from 
$10^{-1}$ (line closest to the square-law structure function 
model) to $10^{1.5}$ (line closest to the simple Kolmogorov 
model), in equal logarithmic steps.}
\label{fig:vela}
\end{figure}

\begin{figure*}[tp]
\centerline{
\psfig{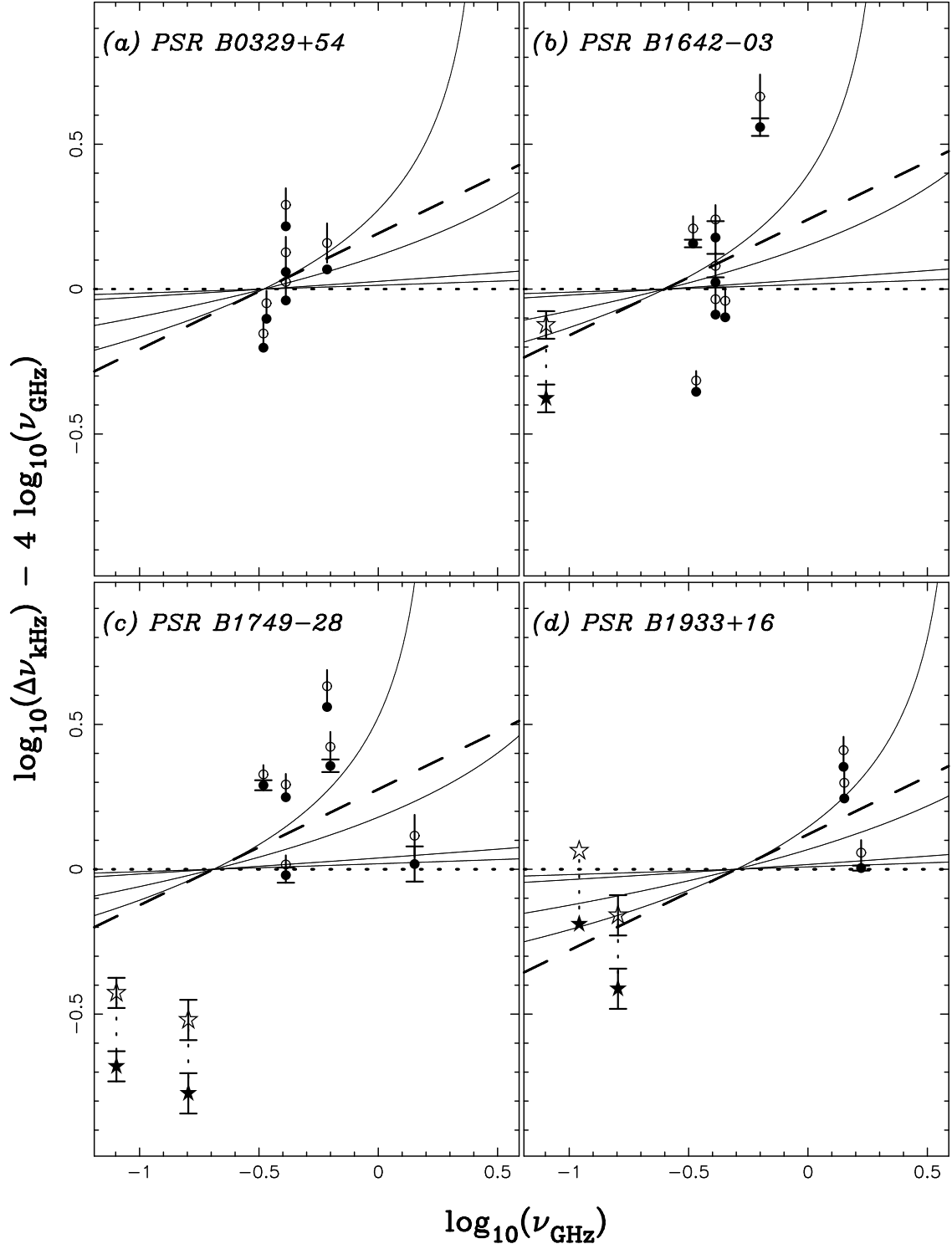}
}
\caption{
Diffractive Decorrelation bandwidth versus frequency for 4 pulsars 
plotted as for Figure~\ref{fig:vela}.  Only theoretical curves
for the $\beta=4$, simple Kolmogorov, and square-law structure
function models are shown.}
\label{fig:other}
\end{figure*}

Decorrelation bandwidth data, collected together by Cordes, Weisberg,
\& Boriakoff (1985; hereafter, CWB), for pulsars PSR B0833$-$45
(Vela), PSR B0329$\-+$54, PSR B1642$\--$03, PSR B1749$-$28, and
PSR B1933$\-+$16 are shown in Figures \ref{fig:vela} and 
\ref{fig:other}.  We have also included recently measured
points by Johnston et al.\ (1998) for the Vela pulsar.  These
are $(\nu, \Delta\nu_d) = (8.4\mbox{GHz}, 13.9\mbox{MHz})$ and
$(13.7\mbox{GHz}, 58.2\mbox{MHz})$.
The data are shown as solid circles
if derived from dynamic spectra and as solid stars if derived
from pulse broadening measurements. The latter become easier to 
estimate at low frequencies, where the resolution bandwidth 
required for the former becomes too narrow for an adequate 
signal-to-noise ratio. The conversion of an estimate
of a pulse broadening time to a decorrelation bandwidth
relies on the uncertainty relation $2 \pi \Delta\nu_d \tau_d = C_1$. 
Unfortunately, the ``constant'' $C_1$ takes on
different values for different geometries and spectral models. 
For the $\beta=4$ model, it is also very weakly dependent on frequency.  
In Table 1 of LR99, we gave numerical values for $C_1$ 
for the three spectral models 
with spherical waves in both a screen 
and an extended medium geometry. For a given geometry,
the smallest $C_1$ is for the Kolmogorov spectrum and 
the largest (by about 50\%) is for the square-law 
structure function. Thus in Figures \ref{fig:vela} 
and \ref{fig:other} we plot, as solid
and open stars, the values converted using $C_1$ for the Kolmogorov
and square-law structure function models, respectively.
We display the predictions for extended medium and note that the screen values
are only about 20\% less for each spectrum.   
Thus in comparing data with the theory for the
$\beta = 4$ and inner-scale models, the data should lie between
the extremes of the open and solid stars,
depending on the outer and inner scales, respectively. 
Note that the value of $C_1$ in equation (7) of the 
Taylor et al.\ (1993) catalog is more than 50\% greater than 
values computed by LR99, since it concerns the mean pulse delay 
rather than the 1/e decay time of the pulse.

%

The second correction to the data concerns the effects of
refractive scintillation at the higher radio frequencies, which
are closer to the transition to weak scintillation.
Gupta et al.\ (1994) gave a heuristic theory of the effect
and expressions for the bias to the diffractive
$\Delta\nu_d$ due to refractive shifts; LR99 discussed the
effect from the work of Codona et al.\ (1986; hereafter, CCFFH). 
In Figures 
\ref{fig:vela} and \ref{fig:other}, we have plotted as open 
circles $\Delta\nu_d$ corrected
for this effect using equation (D6) of Gupta et al.\ (1994), who also
estimated the variability in estimates of
$\Delta\nu_d$ due to the refractive modulation. We 
used this to estimate an error bar on the open circles,
which is typically larger than the error bars quoted by
the original observers, plotted on the solid circles.

In summary, the open circles and their error bars
provide the best direct estimates of $\Delta\nu_d$
and the indirect estimates lie between the 
open and solid stars.

\subsection*{Theory}

The theory of diffractive scintillations has been discussed 
by many authors; we will use the results of section 6 in LR99 
for a point source (spherical waves) either in an extended scattering
medium or with a screen at $z_p$ from the pulsar and $z_o$ from 
the observer (with $z_p+z_o = L$).
Following usual practice, we assume the strong scattering limit,
in which the frequency decorrelation function 
for intensity is the squared magnitude of the diffractive
component of the two-frequency second moment of the field.
The validity of this approximation is examined in Appendix B,
where we find that it introduces a small error for the
$\beta=4$ model, which however is negligible compared to
the errors in the observations.
The decorrelation bandwidth, $\Delta\nu_d$, can be defined
in terms of a normalized bandwidth, $v_d$:
\begin{equation}
\Delta \nu_d  = \frac{v_d \nu}{u^2}
\;\;\;{\rm where}\;\; u^2 = \frac{z_{\rm scatt} c}{2 \pi  \nu s_{0}^2}
\; .
\label{eq:Delta_nu_d}
\end{equation}
Here $\nu$ is the (geometric) mean of the two frequencies,
and the parameter $u$ determines the strength
of scattering; $c$ is the speed of light and $z_{\rm scatt}$ 
is the effective scattering distance.
For a uniform scattering medium, $z_{\rm scatt} = L$, and
for a screen geometry, $z_{\rm scatt} = z_e = z_o z_p/L$.
$s_{0}$ is the field coherence scale where the appropriate phase 
structure function equals unity; for the  
uniform scattering medium, it is defined at the observer,
and for a screen, it is $s_{\rm 0,scr}$ which
is independent of the distances to the pulsar and screen,
and leads to $u_{\rm scr}$ as the appropriate
strength of scattering parameter.

The computations presented in LR99 include a tabulation of the
normalized bandwidth $v_d$ for the various models under discussion, 
and associated plots of the intensity decorrelation function
itself.  For the simple Kolmogorov model, a constant
value for $v_d$ is obtained regardless of the frequency or
distance (0.773 and 0.654  for the extended medium 
and screen geometries, respectively). Thus the frequency
dependence of $\Delta\nu_d$ reflects the frequency dependence
of $s_0$, giving $\Delta\nu_d \propto \nu^{2\beta/(\beta-2)}$.
CWB used this to estimate $\beta$ from observations of
$\Delta\nu_d$.  We now compare the observations
with theoretical scaling laws for the two other spectral models.
There are, however, some complications. 
The quantity $v_d$ is no longer exactly independent
of frequency, and $s_0$ depends on both frequency and 
the outer scale, $L_0$, or the inner scale, $L_i$, that
parameterize the other two models.   

Consider the details for the discontinuity spectrum.
In Figure \ref{fig:vela}, 
we plot $\Delta \nu_d/\nu_{\rm GHz}^4$ versus $\nu_{\rm GHz}$ 
as observed for the Vela pulsar. 
In Figure \ref{fig:vela}a, we have plotted theoretical
curves for $\beta=4$ model with various outer scales.
We use equation (\ref{eq:Delta_nu_d}) for the models
and so need the variation of both $s_0$ and $v_d$ with $\nu$.

%
\begin{figure*}[tb]
\centerline{
\psfig{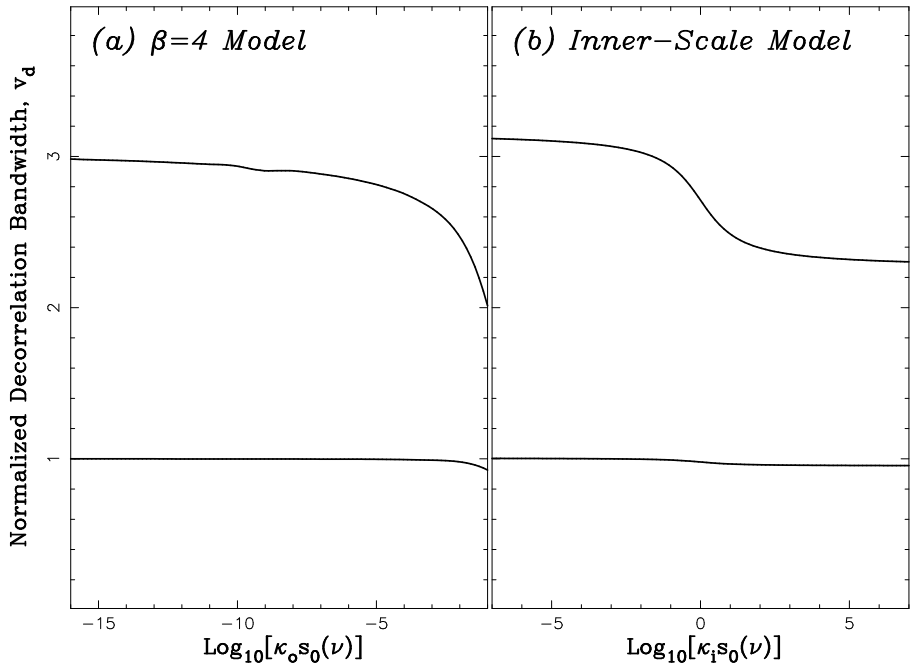}
}
\caption{
Variation of the normalized diffractive decorrelation 
bandwidth, $v_d$. (a) The $\beta=4$ model,
upper curve for an extended medium
geometry with parameter $\kappa_o s_0$  and 
lower curve for a screen geometry versus $\kappa_o s_{0,\rm scr}$.
(b) Same plots for the inner-scale model.}
\label{fig:vd}
\end{figure*}

For a screen, we determine $s_{\rm 0,scr}$ from equation 
(\ref{eq:struc4}) for the $\beta=4$ model,
and eliminate the scattering measure $SM$ using the same equation for
$s_{\rm 0,1GHz}$ at a reference frequency of 1 GHz.
This gives a relation between 
$s_{\rm 0,scr}/s_{\rm 0,1 GHz}$ and $\nu_{\rm GHz}$,
with $L_o/s_{\rm 0,1 GHz}$ as a parameter. 
For an extended scattering medium, we use equation (21) of LR99
to determine $s_0$ at the observer, and the same method as for the screen
to eliminate $SM$.
In Figure \ref{fig:vd}, we show the variation of 
$v_d$ with $L_o/s_{0}$ at a fixed frequency
(computed by Lambert 1998 and described in LR99).
This is combined with the $s_0$--$\nu_{\rm GHz}$ relation to
obtain the relation of $v_d$ to $\nu_{\rm GHz}$.
For the curves in Figure \ref{fig:vela},
$s_{\rm 0,1GHz}$ is determined so that,
for each value of $L_o$, the model $\Delta \nu_d$
fits the measured values near 1GHz. 
For most of the models of interest 
$L_o \gg s_{0}$, and
then the logarithm functions vary much more slowly than the 
$s_0^2$ term in equation (\ref{eq:struc4}). 
In which case there is nearly a 
linear relation $s_0/s_{0,\rm ref} = \nu/\nu_{\rm ref}$.
Furthermore, Figure \ref{fig:vd} shows that 
as $s_0$ changes, the parameter $v_d$ changes very slowly.
Consequently, for a very wide range of the outer scale,
a good approximation is $\Delta \nu_d \propto \nu^4$, which reflects
the fact that the underlying structure
function then approaches a square-law behavior. 

Consider the plot for the Vela pulsar PSR~B0833$-$45
(Figure \ref{fig:vela}) for which the data are most extensive.
Although we have computed curves and data corrections 
for both the screen
and extended scattering geometries, we only show
the extended medium results in Figure \ref{fig:vela}
because the two plots are so similar.  
In the comparison with the screen model,
the star points are lowered about 0.08 vertical units
(20\%), and the $\beta=4$ model curves are very slightly steeper
functions of frequency. However, the scatter among
the observations is greater than the differences in the
models between screen and extended geometries.
In the Vela plot, we see that the $\beta=4$ model agrees
somewhat better with the observations than does the simple
Kolmogorov model.  
This pulsar is known to lie in a highly
scattered region, and hence the presence of discontinuities, as 
incorporated in the $\beta=4$ model, is perhaps reasonable.
However, the conclusion is not strong, since the data are
also reasonably fitted by an inner scale in the range
$100$~km to $500$~km, as can
be found from Figure \ref{fig:vela}b.

For the other pulsars (Figure \ref{fig:other}), 
the data show a stronger frequency dependence than
predicted by both the simple Kolmogorov and $\beta=4$ models. 
However, the inconsistencies between the
measurements at different frequencies are even greater, and
better observations are needed, before such comparisons
could discriminate between the models.  
A recent series of measurements at 327 MHz (Bhat et al.\ 1999a,b,c)
documents the variability of $\Delta\nu_d$, and 
a similar long sequence of such measurements at other frequencies
is needed on the same pulsars, before reliable conclusions can
be reached from the frequency scaling observations.

%
%
\section{REFRACTIVE SCINTILLATION INDEX}
\label{sec:mr}

Slow variations in the flux of pulsars are caused by
refractive interstellar scintillation (RISS),
and can be characterized by the
rms deviation in flux density normalized by its mean,
or scintillation index, $m_R$.
RISS is due to
inhomogeneities in the interstellar electron density on
scales much larger than those responsible for diffractive
scintillation, as characterized by the decorrelation bandwidth.
There are now measurements of both phenomena on a substantial number
of pulsars, from which we can constrain the density spectrum on
each line of sight. The relation of the refractive scintillation 
index, $m_R$, to the normalized diffractive decorrelation bandwidth, 
$\Delta\nu_d/\nu$, depends on the ratio of the power in the density 
spectrum at the large refractive scales ($10^{11}$~m to $10^{12}$~m)
to the power at the smaller diffractive scales ($10^7$~m to $10^8$~m). 


We compare pulsar measurements gathered from the literature
with theoretical predictions by plotting
$m_R$ versus $\Delta\nu_d/\nu$
(cf.\ Rickett \& Lyne 1990 and Gupta et al.\ 1993). Related
tests have been made by Armstrong et al.\ (1995), Bhat et 
al.\ (1999a) and Smirnova, Shishov, \& Stinebring (1998). 
$m_R$ measurements at 610 MHz are listed in 
Table \ref{table:mr610} and include the results of long-term 
monitoring observations by Smirnova et al.\ (1998),
and older measurements near 100~MHz are listed in Table 
\ref{table:mr100}. 
The tables also include the decorrelation bandwidth
obtained from diffractive scintillation, typically observed at a
different radio frequency, scaled to the observing frequency for $m_R$.
We used the Kolmogorov scaling law $\Delta\nu_d \propto \nu^{4.4}$, and
we note that the minor differences that result from changes in the 
scaling law for other spectral models are insignificant in the 
comparison.  The 100~MHz data consist of $m_R$ measurements at 
73.8~MHz, 81.5~MHz, and 156~MHz.  
The pulsars observed at 610 MHz are primarily located at distances
$\simgreat 1$~kpc, whereas the pulsars observed at 100 MHz
are nearer at distances $\simless 1$~kpc.  
Figures \ref{fig:mr4} and \ref{fig:mrin} show
plots of the theoretical and measured refractive scintillation
index versus the normalized decorrelation bandwidth
in separate panels for the two sets of measurements. 
Theoretical curves are also shown in the Figures
for different spectral models in both the screen and extended medium 
geometries.  The dashed line corresponds to the simple Kolmogorov model
($\alpha=5/3$), and the solid lines in Figures \ref{fig:mr4}
and \ref{fig:mrin}  correspond to the $\beta=4$ and inner-scale 
models, respectively.  

In section \ref{sec:decbw} we noted
that variable refraction tends to decrease the measured 
decorrelation bandwidth and used an expression from
Gupta et al.\ (1994) to apply a nominal correction
to measured values.  In the table of data from near 100 MHz,
we have made use of the recently published measurements of 
Bhat et al.\ (1999a). They monitored the apparent decorrelation 
bandwidth near 327 MHz for 20 nearby pulsars, many of which
have  $m_R$ measurements in Table \ref{table:mr100}. They found
the bandwidth to vary by factors 3-5 and discussed the influence
of varying refraction as a refractive bias. 
From their measurements, they derived
a corrected decorrelation bandwidth, which we have scaled
(using the $\nu^{4.4}$ scaling law) 
to the frequency at which $m_R$ was observed.
These gave values typically 2--5 times bigger than
obtained from earlier measurements (e.g.\ Cordes 1986).
The pulsar B0809+74 was not observed by Bhat et al.\ (1999a);
however, the recent weak scintillation observations of this
pulsar by Rickett et al. (1999) similarly suggests that
earlier decorrelation bandwidth measurements
overestimated the strength of scattering.
The effect of these changes is to shift the plots
to the right in Figures \ref{fig:mr4} and \ref{fig:mrin} 
by about half a decade, compared to Figure 5
of Gupta et al.\ (1993).  There is only one
pulsar (B0329+54) in Table \ref{table:mr610}
common to the Bhat et al.\ (1999a) observations, and 
its decorrelation bandwidth has also 
been corrected for the refractive bias. Since the pulsars
observed at 610 MHz were mostly more heavily scattered, the
refractive bias correction is substantially smaller
(see Gupta et al.\ 1994) and has been ignored.
 
\subsection*{Theory}

The theory of refractive scintillations has been known since the 
1970s and was applied to pulsar flux variations by 
Rickett et al.\ (1984).
For example, Prokhorov et al.\ (1975) described 
how the modulation index for intensity can exceed
unity in strong scattering in their equations (4.53) \it et. seq\rm.
We use the notation of Coles et al.\ (1987) and confine our
discussion to spherical wave sources, propagating
in a scattering plasma which is either concentrated in a screen
or extended uniformly between source and observer.
The ``low-frequency'' approximation for the intensity
covariance function is given by equation (10) of Coles et al.\ (1987),
from which we obtain the normalized refractive variance, $m_R^2$,
by setting the spatial offset equal to zero:
\begin{eqnarray}
m_R^2 & = & 
\frac{16 \pi^2 r_{\rm e}^2 c^2 L}{\nu^2}
\int_0^1 \, \int_0^{\infty}
P_{N_e}(\kappa, q_z=0;z) \times \nonumber \\
& & \exp\left\{
- L \int_0^1 D_{\phi}'\left[\kappa r_{{\rm F},L}^2 h(x,y)\right] dy 
\right\} \times 
\nonumber \\
& &
\sin^2\left[ 0.5 \kappa^2 r_{{\rm F},L}^2 x(1-x)\right]
\, \kappa d\kappa \, dx
\, \mbox{,}
\label{eq:crextmed}
\end{eqnarray}
where:
\begin{eqnarray}
h(x,y) & = & {\rm min}\left[y(1-x), x(1-y) \right]  \\
r_{{\rm F},L}& = & \sqrt{Lc/2 \pi \nu} \mbox{.}  \nonumber
\label{eq:h}
\end{eqnarray}
Here the transverse
wavenumber is $\kappa = \sqrt{q_x^2 + q_y^2}$, and
$x=z/L$, where $z$ is the distance along the line of
sight measured from the source.  For the screen geometry, 
both the density spectrum and the gradient in the 
phase structure function $D_{\phi}'$ are concentrated
in a thin layer of thickness $\Delta z$ at distance $z_p$
from the source and $z_o$ from the observer.
Thus the $x$ and $y$ integrations in equation (\ref{eq:crextmed})
give $x=y=z_p/L$, and we find it useful 
to define an equivalent Fresnel scale $r_{\rm F,scr}$:
\begin{equation}
r_{\rm F,scr} =  \sqrt{z_{\rm e}c/2 \pi \nu} \;\; 
\mbox{where} \;\; z_{\rm e}=z_p z_o/L
\, \mbox{.}
\label{eq:rfscr}
\end{equation}
The $\kappa$ integration  in equation (\ref{eq:crextmed})
involves the product of the density spectrum, decreasing
as a steep power of $\kappa$, times the high-pass Fresnel filter
$\propto \kappa^4$ up to $r_{\rm F,scr}^{-1}$, times
the low-pass exponential term that cuts off
wavenumbers above $s_R^{-1}$, where  $s_R = u_{\rm scr} r_{\rm F,scr}$
is the radius of the scattering disc. 
Hence in strong scattering ($u_{\rm scr} > 1$)
we only consider the $\kappa^4$ part of the Fresnel filter.
Consequently, the integration depends on the ratio
$L_o/s_R$ or $L_i/s_R$. 
For an extended scattering medium the line-of-sight integration
softens the exponential cut-off in the $\kappa$
integration, but the basic relationships remain the same.
As others have noted the level of refractive scintillation
is greater in an extended medium than in a screen with the
same observed diffractive scintillation.

\begin{figure*}[tb]
\centerline{
\psfig{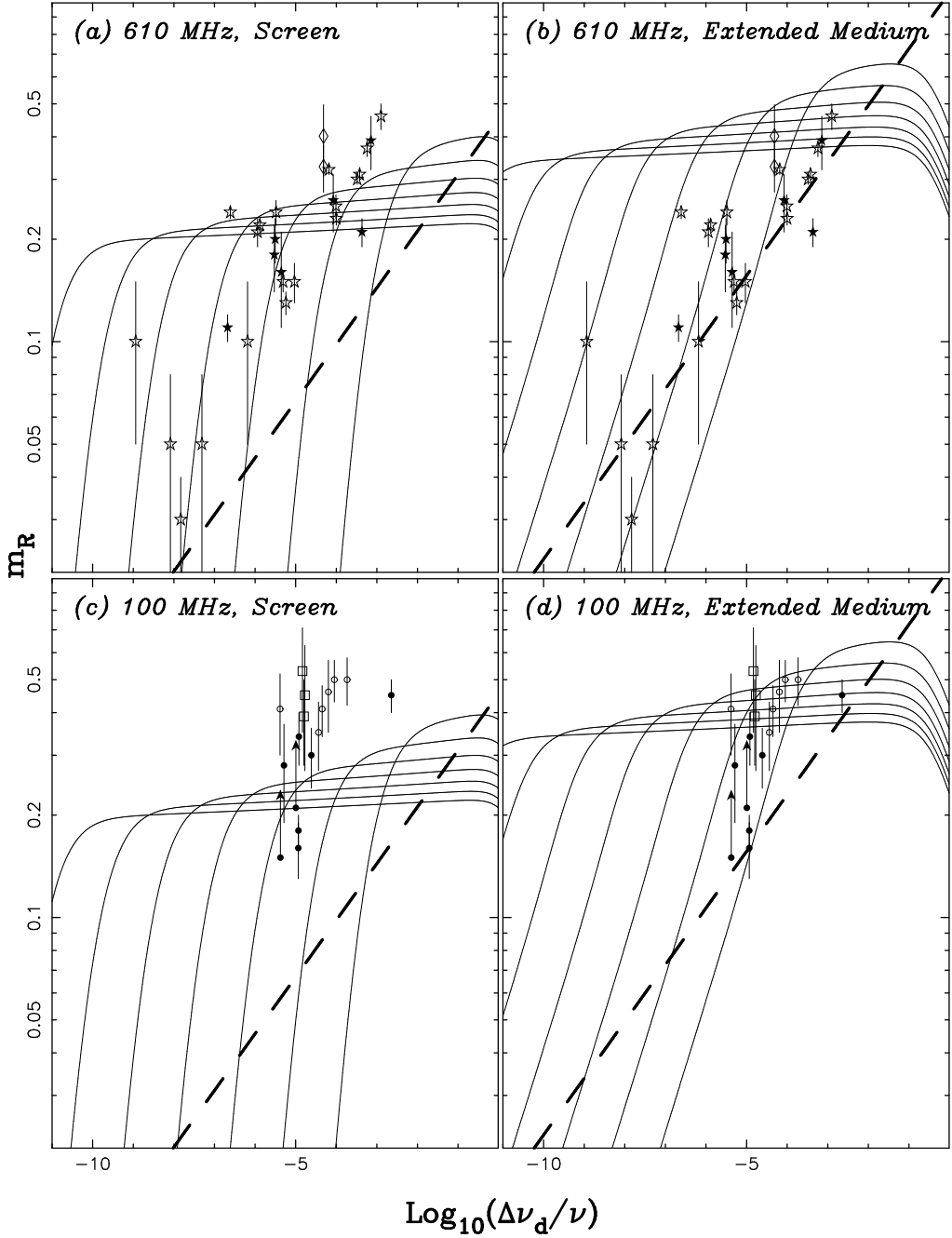}
}
\caption{
Theoretical and measured refractive scintillation
indices, $m_R$, versus the normalized diffractive decorrelation 
bandwidth, $\Delta\nu_d/\nu$. (a) and (b) show measurements at
610~MHz; (c) and (d) correspond to measurements near 100~MHz. 
The solid curves give the theoretical results for the 
$\beta=4$ model with a range of outer scales from  
$10^{10}$~m to $10^{16}$~m, going from right to left in 
equal logarithmic steps.
The dashed line corresponds to the simple
Kolmogorov model in the strong scattering limit.
Panels (a) and (c) give theory for 
a screen geometry; panels (b) and (d) for a uniform scattering medium.}
\label{fig:mr4}
\end{figure*}

\begin{figure*}[tb]
\centerline{
\psfig{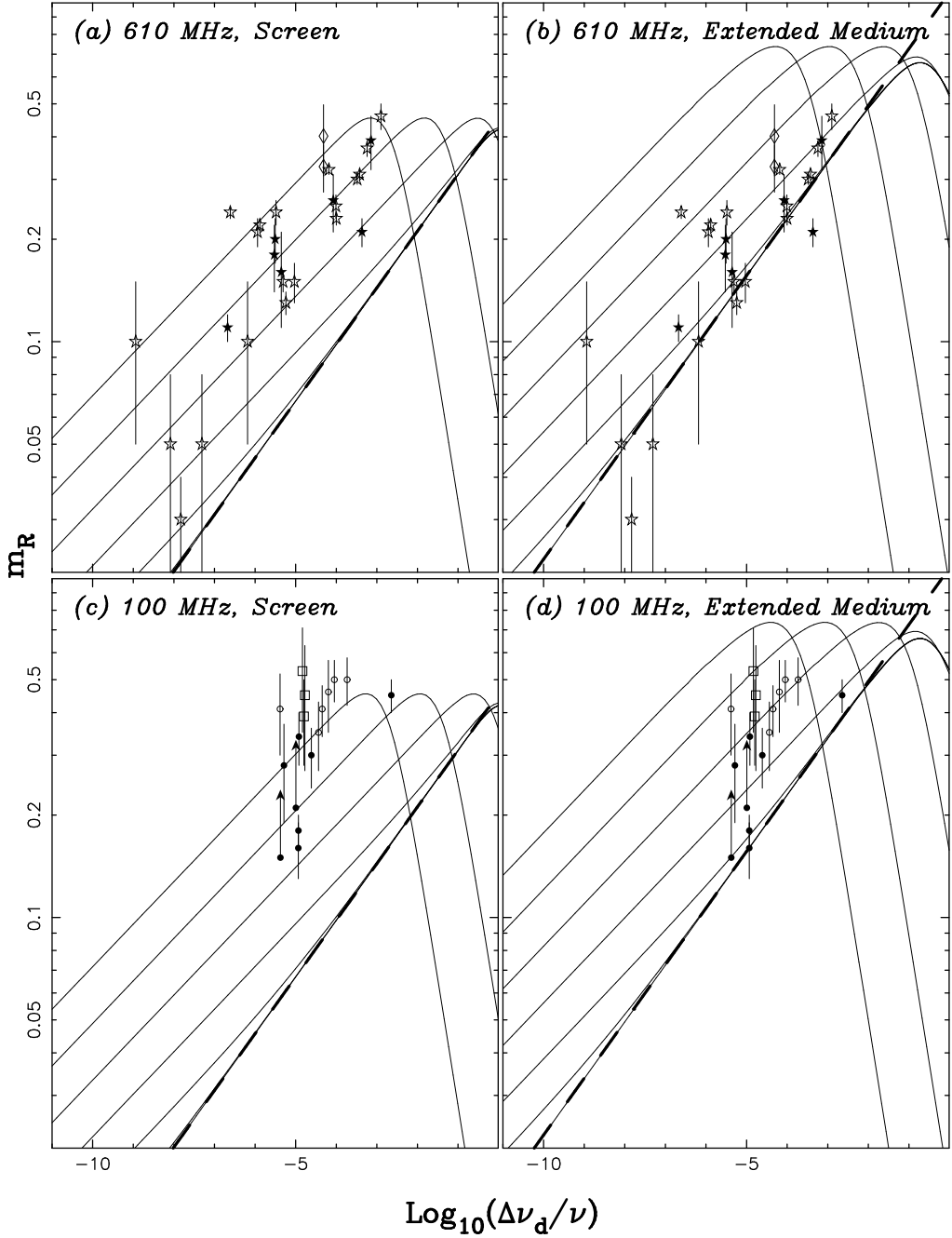}
}
\caption{
Theoretical and measured refractive scintillation
indices, $m_R$, versus the normalized diffractive decorrelation 
bandwidth, $\Delta\nu_d/\nu$. (a) and (b) show measurements at
610~MHz; (c) and (d) correspond to measurements near 100~MHz. 
The solid curves give the theoretical results for the 
inner-scale model with a range of inner scales from 
0 (corresponding to the simple Kolmogorov model) to
The dashed line corresponds to the simple
Kolmogorov model in the strong scattering limit.
Panels (a) and (c)---screen geometry; 
panels (b) and (d)---uniform scattering medium.}
\label{fig:mrin}
\end{figure*}

The canonical spectral model in ISS studies of pulsars
has been the simple Kolmogorov spectrum
$P_{N_e}(q,z) = C_{N_e}^2(z) q^{-11/3}$. By fitting
this model to diffractive scintillation observations,
observers have estimated the scattering measure,
$SM = \int_0^L C_{N_e}^2(z) dz$, toward many pulsars.
When divided by the pulsar distance, this gives
a line-of-sight average of $C_{N_e}^2(z)$, which is found
to vary greatly from one direction to another 
and to increase dramatically for distant pulsars seen at low
galactic latitudes (e.g.\ CWB).  
Taylor \& Cordes (1993) have developed a smoothed model for the
Galactic plasma density distribution, which includes
enhanced scattering in spiral arms and toward the Galactic center.
However, there are also large random variations on much finer
spatial scales,
which would produce a scatter in a plot of $m_R$ against distance
of dispersion measure.
For a simple power-law spectrum both $m_R$ and 
$\Delta\nu_d/\nu$ depend on $SM$ through
a single strength of scattering parameter. Thus for the
Kolmogorov spectrum the variation of $m_R$ with $\Delta\nu_d/\nu$ 
is independent of distance or frequency and is given by a single 
dashed line in each plot in Figures \ref{fig:mr4} and \ref{fig:mrin}. 
If the scattering medium is uniform, the strength of scattering
is $u = r_{{\rm F},L}/s_0$. For the screen it becomes
$u_{\rm scr} = r_{\rm F,scr}/s_{\rm 0,scr}$, and
in that case, the dashed theoretical line is independent
of the location of the screen between the source and the observer.
This point is substantiated in Appendix~A, where the
theory is laid out in more detail. According to this screen model,
along each line of sight there is a scattering layer with a certain
$SM$, which determines both $m_R$ and $\Delta\nu_d/\nu$,
but $SM$ is not necessarily related to the pulsar distance. 
The theoretical curve with $SM$ as the variable is
independent of where the layer is located along the line of sight.

For the $\beta=4$ and inner-scale models, the theoretical
curves for $m_R$ versus $\Delta\nu_d/\nu$
depend on the extra parameter $L_o$ or $L_i$, respectively.
Details of the theory are given in Appendix~A,
where the relevant parameters are shown to be
$L_o/s_{\rm 0,scr}$ and $L_i/s_{\rm 0,scr}$.
Since $s_{\rm 0,scr}$ depends on frequency,
$SM$, and distance, then $m_R$ also
depends somewhat on frequency and distance.
In order to fix the frequency dependence
of the theoretical $m_R$ values, we have 
separated the measurements into two groups (at 610~MHz and
near 100~MHz).  The distance dependence
is dealt with by assuming that $SM \propto$ distance.
Whereas this is clearly appropriate for the extended
scattering medium, it is less clear for the screen model,
since the screen model supposes that $SM$ is not necessarily 
related to distance. However, it is reasonable that, if
on a long line of sight there is a single region that
dominates the scattering, its $SM$ value will
statistically increase with line-of-sight distance.
Indeed the experimentally derived scattering measure
increases faster with distance than if the medium
were uniform (see Figure 1 of Cordes et al.\ (1991).

%

In Figure \ref{fig:mr4} we show the theoretical 
curves for $m_R$ in the $\beta=4$ model, which  
are relatively flat where $L_o > s_R$. As
$u$ increases ($\Delta\nu_d/\nu$ decreases),
$s_R$ increases, and when $s_R > L_o$ the scintillations
are suppressed.  In a screen
this occurs when $\Delta\nu_d/\nu \simless (r_{\rm F,scr}/L_o)^2$.

Inspection of Figure \ref{fig:mr4} shows that
the measured $m_R$ values at 100~MHz are 
above the prediction of the simple Kolmogorov model for the 
extended medium geometry.  These measurements can,
however, be modeled with the $\beta=4$ spectrum with 
suitable specific choices of outer scale, $L_o$.  
The screen geometry is inconsistent with 
the measurements at 100~MHz for both the simple Kolmogorov and 
$\beta=4$ models.  For the 610~MHz data, the measured $m_R$ 
values are mostly in agreement with the prediction of the simple 
Kolmogorov model for the extended medium geometry; however, they 
lie below the predictions of the $\beta=4$ model.  For the
screen geometry, the measured $m_R$ values are above the 
simple Kolmogorov model curve.  However, for the $\beta=4$ 
model, though less convincingly, an agreement with the 
measurements can be found by  suitable choices of the outer 
scale, $L_o$.  The most striking feature of the Figure
is the good agreement with the uniform extended Kolmogorov
model for most of the pulsars.  
This result, relies most heavily on the excellent
data from Smirnova et al.\ (1998), who also note this
agreement.  We are persuaded by this Figure that the $\beta=4$
model cannot be viewed as a global alternative to the Kolmogorov
spectrum. We also conclude that for the medium-distance pulsars
measured at 610~MHz, $m_R$ does not agree with the screen
geometry.  It appears that even if the scattering medium is not uniform
on scales of a few kpc, there is not a single region that
dominates the scattering.  Computations of $m_R$
due to a patchy distribution along the line of sight could
test what distribution would start to approximate the
uniformly extended medium.

%

Turning to a comparison with the alternative inner-scale spectrum in 
Figure \ref{fig:mrin}, the 100~MHz observations
could be explained by very large values of $L_i$
($10^{10}$m to $10^{11}$m) in a screen
configuration or by more modest, but still large, values
of $L_i$ ($10^8$m) in an extended medium.  
The latter was essentially the proposal made 
by Coles et al.\ (1987) for the nearby lines of sight.
The 610 MHz points are relatively lower than those at 100 MHz.
For an inner scale spectrum in a screen, large 
values of $L_i$ ($10^8$m to $10^{10}$m) would be necessary;
and in an extended medium 60\% of the points
lie near the simple Kolmogorov model, with the rest requiring
inner scale values ($10^7$m to $10^8$m).
The 610 MHz data in our analysis come from
RISS observations of 21 pulsars made by Stinebring and are
described in more detail by Smirnova et al.\ (1998), 
who also compare the results with 
various models for the spectrum and spatial distribution 
of the electron density. They note that 4 pulsars which 
are seen through known HII regions or supernovae remnants
show relatively elevated values for $m_R$, and find 3 other
pulsars with similar behavior.  These are the points which lie
above the Kolmogorov line in Figure \ref{fig:mrin}b.
Their interpretation is that for these objects 
the scattering is concentrated into regions either near
the pulsar or near the Earth, and that these regions
are characterized by an inner scale near $3\times 10^8$m.
Their spectrum model is very similar to our inner-scale model, 
except that the cut-off is characterized
by a steep power-law rather than by an exponential function.
We find a somewhat larger numerical value for the inner
scale required to match those objects.  We also note
that the theoretical $m_R$ values are consistently 
higher for an extended medium, and so require 
a more modest inner scale.

However, an important result of our analysis
is an alternative explanation for the relatively high $m_R$
values seen for some pulsars. We suggest that
pulsar lines of sight can pass through discrete clouds
with increased plasma density on large scales, 
that steepen the low-wavenumber spectrum as opposed to cutting 
off the high-wavenumbers. A discontinuity spectrum
($\beta=4$) is one way that the spectrum can be steepened,
but smoother structures on scales larger than
$10^{11}$m ($\sim 1$ AU) would also boost the low-wavenumbers
and could cause similar enhancements of refractive
compared to diffractive scintillations.  
Such enhancements are likely to be associated with HII
regions or supernova remnants. A similar
idea was also discussed by Lestrade, Rickett, \& Cognard (1998)
in the context of extreme scattering events in
pulsar timing measurements.  To take this idea further,
a composite between the Kolmogorov and 
the $\beta=4$ or steeper spectra should be investigated, particularly
one in which the line of sight is not uniformly weighted.
Smirnova et al.\ (1998) also note that the strength of 
scattering increases very
much faster with distance and dispersion measure than if the medium
were statistically uniform. The distant pulsars in their sample
are mostly observed at low galactic latitudes and so are
subject to the enhanced density and turbulence described 
by the Taylor \& Cordes (1993) model. It appears that
most distant pulsars follow the uniform extended Kolmogorov
model with inner scale smaller than $10^{6}$m.  Thus although
these lines of sight are subject to enhanced scattering 
in the inner galactic plane, the spectrum effectively follows
the Kolmogorov law and the plasma is dispersed enough to 
approximate a uniform scattering medium. 
There are presumably discrete refracting structures
present along these lines of sight, but their contribution to
the density spectrum is masked by
the higher densities of the inner Galaxy, which still follow an
apparently turbulent spectral form.

\begin{deluxetable}{llr@{.}lr@{.}lr@{.}lr@{.}lr@{.}l}
\renewcommand{\arraystretch}{0.8}
\tablecaption{Refractive modulation index data at 610~MHz
from the literature. 
\label{table:mr610}
}
\footnotesize
\tablecolumns{12}
\tablewidth{0pt}
\tablehead{
\colhead{Reference} &
\colhead{Pulsar} &
\multicolumn{2}{c}{Distance} &
\multicolumn{2}{c}{$\nu$} &
\multicolumn{2}{c}{$\Delta\nu$} &
\multicolumn{2}{c}{$m_r$} &
\multicolumn{2}{c}{$\delta m_r$} 
\\
\colhead{} &
\colhead{} &
\multicolumn{2}{c}{kpc} &
\multicolumn{2}{c}{MHz} &
\multicolumn{2}{c}{MHz} &
\colhead{} &
\colhead{} &
\colhead{} &
\colhead{}
}
\startdata
Rickett \& Lyne (1990)

& B0531+21   & 2&0  & 610&0 & 2&96$\times 10^{-2}$ 
& 0&327 & 0&052 \nl

& B0531+21   & 2&0  & 610&0 & 2&96$\times 10^{-2}$
& 0&402 & 0&095 \nl 

Kaspi \& Stinebring (1992) 

& B0329+54   & 1&4  & 610&0 & 0&44
& 0&39  & 0&07  \nl

& B0833$-$45 & 0&55 & 610&0 & 1&29$\times 10^{-4}$
& 0&11  & 0&01  \nl

& B1749$-$28 & 1&2  & 610&0 & 5&13$\times 10^{-2}$
& 0&26  & 0&05  \nl

& B1911$-$04 & 2&29 & 610&0 & 1&91$\times 10^{-3}$
& 0&20 & 0&03  \nl

& B1933+16   & 7&8  & 610&0 & 1&82$\times 10^{-3}$
& 0&18  & 0&04  \nl

& B2111+46   & 5&22 & 610&0 & 2&69$\times 10^{-3}$
& 0&16  & 0&05  \nl

& B2217+47   & 2&31 & 610&0 & 0&26
& 0&21  & 0&02  \nl

Stinebring  et al.\ (1999)

& B0136$+$57 & 2&9 & 610&0 & 5&7$\times 10^{-3}$
& 0&15 & 0&02 \nl

& B0329$+$54 & 1&4 & 610&0 & 0&350
& 0&37 & 0&02 \nl

& B0525$+$21 & 2&3 & 610&0 & 0&30
& 0&31 & 0&01 \nl

& B0531$+$21 & 2&0 & 610&0 & 4&0$\times 10^{-2}$
& 0&32 & 0&01 \nl

& B0736$-$40 & 2&1 & 610&0 & 9&0$\times 10^{-6}$
& 0&03 & 0&01 \nl

& B0740$-$28 & 1&9 & 610&0 & 3&5$\times 10^{-3}$
& 0&13 & 0&01 \nl

& B0818$-$13 & 2&5 & 610&0 & 0&061
& 0&23 & 0&01 \nl

& B0833$-$45 & 0&5 & 610&0 & 1&50$\times 10^{-4}$
& 0&24 & 0&01 \nl

& B0835$-$41 & 4&2 & 610&0 & 7&0$\times 10^{-4}$
& 0&21 & 0&02 \nl

& B1641$-$45 & 4&6 & 610&0 & 7&0$\times 10^{-7}$
& $\sim$0&1 & 0&05 \nl

& B1642$-$03 & 0&5 & 610&0 & 0&770 
& 0&46 & 0&04 \nl

& B1749$-$28 & 1&5 & 610&0 & 6&0$\times 10^{-2}$
& 0&25 & 0&02 \nl

& B1818$-$04 & 1&6 & 610&0 & 4&0$\times 10^{-4}$
& $\sim$0&1 & 0&05 \nl

& B1859$-$03 & 8&1 & 610&0 & 5&0$\times 10^{-6}$
& $\sim$0&05 & 0&03 \nl

& B1911$-$04 & 3&2 & 610&0 & 8&0$\times 10^{-4}$
& 0&22 & 0&01 \nl

& B1933$+$16 & 3&5 & 610&0 & 2&0$\times 10^{-3}$
& 0&24 & 0&02 \nl

& B1946$+$35 & 7&9 & 610&0 & 3&0$\times 10^{-5}$
& $\sim$0&05 & 0&03 \nl

& B2111$+$46 & 5&0 & 610&0 & 3&0$\times 10^{-3}$
& 0&15 & 0&01 \nl

& B2217$+$47 & 2&5 & 610&0 & 0&20
& 0&30 & 0&01 \nl
\enddata
\end{deluxetable}

%
%

\begin{deluxetable}{llr@{.}lr@{.}lr@{.}lr@{.}lr@{.}l}
\renewcommand{\arraystretch}{0.8}
\tablecaption{Refractive modulation index data at 100~MHz
from the literature. 
\label{table:mr100}
}
\footnotesize
\tablecolumns{12}
\tablewidth{0pt}
\tablehead{
\colhead{Reference} &
\colhead{Pulsar} &
\multicolumn{2}{c}{Distance} &
\multicolumn{2}{c}{$\nu$} &
\multicolumn{2}{c}{$\Delta\nu$} &
\multicolumn{2}{c}{$m_r$} &
\multicolumn{2}{c}{$\delta m_r$} 
\\
\colhead{} &
\colhead{} &
\multicolumn{2}{c}{kpc} &
\multicolumn{2}{c}{MHz} &
\multicolumn{2}{c}{MHz} &
\colhead{} &
\colhead{} &
\colhead{} &
\colhead{}
}

\startdata

Cole et al.\ (1970)  

& B0809+74 & 0&31 & 81&5  & 1&38$\times 10^{-3}$
& 0&45 & 0&18    \nl

& B0834+06 & 0&72 & 81&5  & 1&3$\times 10^{-3}$
& 0&39 & 0&11    \nl

& B1919+21 & 0&66 & 81&5  & 1&2$\times 10^{-3}$
& 0&53 & 0&18    \nl 

Helfand et al.\ (1977)  

& B0329+54 & 1&4 & 156&0 & 5&7$\times 10^{-3}$
& 0&35 & 0&08    \nl 

& B0823+26 & 0&38 & 156&0 & 1&0$\times 10^{-2}$
& 0&46 & 0&11    \nl

& B1133+16 & 0&27 & 156&0 & 2&9$\times 10^{-2}$
& 0&50 & 0&08    \nl

& B1508+55 & 1&93 & 156&0 & 7&0$\times 10^{-3}$
& 0&41 & 0&07    \nl

& B1919+21 & 0&66 & 156&0 & 1&4$\times 10^{-2}$
& 0&50 & 0&07    \nl

& B2217+47 & 2&31 & 156&0 & 6&45$\times 10^{-4}$
& 0&41 & 0&11    \nl

Gupta et al.\ (1993) 

& B0329+54 & 1&4  & 73&8  & 3&1$\times 10^{-4}$
& $>$0&15 & \multicolumn{2}{c}{\nodata} \nl 

& B0809+74 & 0&31 & 73&8  & 8&90$\times 10^{-4}$
& 0&34 & 0&06    \nl

& B0834+06 & 0&72 & 73&8  & 8&6$\times 10^{-4}$
& 0&16 & 0&03    \nl

& B0950+08 & 0&12 & 73&8  & 0&17
& 0&45 & 0&05    \nl

& B1133+16 & 0&27 & 73&8  & 1&6$\times 10^{-3}$
& 0&18 & 0&02    \nl

& B1237+25 & 0&56 & 73&8  & 1&8$\times 10^{-3}$
& 0&30 & 0&06    \nl

& B1508+55 & 1&93 & 73&8  & 3&8$\times 10^{-4}$
& 0&28 & 0&09    \nl

& B1919+21 & 0&66 & 73&8  & 7&5$\times 10^{-4}$
& $>$0&21 & \multicolumn{2}{c}{\nodata} \nl

\enddata
\end{deluxetable}

%
%

\section{SUMMARY AND CONCLUSION}
\label{sec:conclusion}

In this paper, the theory of the $\beta=4$ model for the
electron density spectrum was derived for
discontinuous density structures 
and compared with pulsar observations.  A new feature of our analysis is
the inclusion of an ``outer scale'' needed
in any realistic model. The model is characterized by
an effective exponent $\alpha(s)$
of the structure function, which remains between 1.95 and 1.6
over a very wide range of $L_o$ values (cf.\ Figure \ref{fig:slope}).  
This at first seems a promising explanation for the spread in the 
estimates of $\alpha$ derived from VLBI observations of the
angular broadening profile, as observed on heavily
scattered lines of sight. 

As discussed in section \ref{sec:decbw},
from Figures \ref{fig:vela} and \ref{fig:other},
we find that the $\beta=4$
model provides a somewhat better agreement with the
measurements of the diffractive decorrelation bandwidth versus
frequency for pulsar PSR B0833$-$45 (Vela) than does the simple
Kolmogorov model.  This might arise from refractive scattering
effects caused in the supernova remnant associated with the
Vela pulsar.  Four other pulsars with
decorrelation bandwidths measured against frequency show an
appreciably stronger frequency dependence
than the predictions of both the simple Kolmogorov and
$\beta=4$ models.  However, there are substantial
inconsistencies among the measurements and better observations
are clearly needed,
especially in view of the variability in $\Delta \nu_d$
documented by Bhat at al. (1999c).

The predictions of the $\beta=4$ model for the variation of
the refractive scintillation index with the diffractive
decorrelation bandwidth are in partial agreement with the
observations.  As discussed in section \ref{sec:mr}, 
the values of $m_R$ measured near 100~MHz are above the
prediction of the simple Kolmogorov model and 
had previously been explained as the effect of an inner scale,
substantially larger 
than the values invoked by Spangler \& Gwinn 1990.  
However, the measurements are also consistent 
with the $\beta=4$ model with suitable choices
of the outer scale.  For the 610~MHz data,
most of the measured $m_R$ values are
in good agreement with the prediction of the simple Kolmogorov
model for the extended medium, and they lie below the
curves of the $\beta=4$ model.  However, there are a significant
number of good-quality observations, which lie somewhat above
the Kolmogorov line.  We suggest an alternative to a large
inner scale on those lines of sight, that they pass through 
regions of enhanced density, which cause enhanced refractive scattering;
these regions must have less small-scale substructure than in a 
turbulent medium and could include discontinuities.

Based on the above considerations, we reject the $\beta=4$
model as a universal spectral model for the interstellar
electron density fluctuations. 
The corollary is to strengthen the evidence for 
the Kolmogorov density spectrum, which in turn 
suggests a turbulent process in the 
interstellar plasma. However, the simple Kolmogorov 
spectrum is not a universal model either, 
since it disagrees with several of the $m_R$ observations.
Since the $\beta=4$ model provides reasonable agreement 
for many of these discrepant observations, we propose
that enhancements in the large scale part of the spectrum
(which need not be described by discontinuities in density)
occur on these lines of sight. With such enhancements
causing the increase in refractive scintillation, there
is no need to invoke the relatively large inner scales
proposed by Coles et al. (1987). As proposed by 
Spangler \& Gwinn (1990), a relatively small inner scale
is then likely, controlled by the ion inertial length 
or Larmor radius.

\begin{table*}[tb]
\renewcommand{\arraystretch}{0.8}
\caption{Symbols Used}
\begin{center}

\begin{tabular}{ ll }
\hline 
		&		\\
Symbols & Definition  \\
		&	\\
\hline 

$\alpha$, $\alpha(s)$	& Exponent of separation $s$ in structure function \\
$\beta$		& Exponent of wavenumber in density power spectrum \\
$\mbox{\boldmath $\beta$}$, $\mbox{\boldmath $\beta$}'$ & Vector spatial offsets in Appendix B \\
$c$	& speed of light \\
$C_1$	& Constant in uncertainty relation, $2\pi\Delta \nu_d\tau_d = C_1$ \\
$C_{N_e}^2$		& Coefficient in electron-density wavenumber spectrum  (m$^{-20/3}$) \\
$D_{\phi}$,$D_{\phi}'$		& Wave/phase structure function and its longitudinal gradient Equation (6)\\
$D_{\rm S}$	& Wave/phase structure function for a point (spherical wave) source \\
$\Delta z$	& Screen or layer thickness	\\
$\Delta \nu_d$	& Diffractive Decorrelation Bandwidth	\\
$\epsilon$	& Fractional frequency difference $|k_1-k_2|/(k_1+k_2)$ \\
$f(r), f_0$	& Radial profile function for electron density, interior density \\		
$F(q)$	& Three-dimesnsional Fourier Transform of $f(r)$ Equation (3)\\
$g(\eta), g_1(\eta)$	& Intermediate functions of $\eta = \kappa^2$ \\
$J_0$	& Zero-order Bessel function of the first kind	\\
$\kappa$	& Transverse (2-dimensional) wavenumber \\
$\kappa_i,\kappa_o$	& High and low wavenumber cutoffs Equation (1)\\ 
$k=2\pi/\lambda, k_1, k_2$	& Radio wavenumbers \\
$k_m$	& Geometric mean of $k_1,k_2$	\\
$L$	& Pulsar distance \\
$L_i,L_o$	& Inner and Outer scales \\
$\lambda$	& Radio wavelength \\
$m_R$	& Scintillation index (rms/mean) for refractive scintillation Equation (10)\\
$m_{\rm int}$	& Scintillation index due to scales intermediate between diffractive and refractive\\
$n_0$	& Spatial number density of discontinuous objects \\
$\nu$	& Radio frequency \\
$P_{N_e}(q)$	& Wavenumber spectrum for the electron density Equation (1) \\
$q$		& Three-dimensional wavenumber 	 \\
$r_e$		& Classical electron radius 	\\
$r_{{\rm F},L}$	& Fresnel scale ($\sqrt{L/k}$) Equation (11) \\
$r_{\rm F,scr}$	& Fresnel scale for a screen ($\sqrt{z_e/k}$) Equation (12)\\
$s$	& Transverse separation \\
$SM$	& Scattering measure = Line of sight integral of $C_{N_e}^2$ \\
$s_0$	& Field coherence scale \\
$s_R$	& Refractive scintillation scale \\
$\tau_d$	& Pulse decay time to 1/e \\
$u$	& Strength of scattering parameter ($=r_{{\rm F},L}/s_0$) Equation (9) \\
$v_d$	& Normalized decorrelation bandwidth Equation (9) \\
$V_4$	& Sum of wave structure functions Equation (B1)\\
$x, y$	& Normalized distances \\
$z, z_p$	& Distance from pulsar along line of sight toward observer \\
$z_o$	& Observer-screen distance \\
$z_e$	& Effective screen observer-distance $=z_o z_p/L$ \\
$\zeta$	& $(s_R \kappa_o)^2$ \\

\hline 

\end{tabular}
\end{center}
\end{table*}

It appears that different spectral models need to be
considered for different lines of sight.
A widely distributed turbulent
plasma with occasional large ionized structures that increase the
effective average power density, $P_{N_e}$, at low
wavenumbers (large scales: $10^{11}$~m to $10^{14}$~m)  is
thus a model that needs further formal investigation.  
Very similar conclusions have been reached by
Lestrade et al.\ (1998) and by Bhat et al.\ (1999b,c).
This model could also explain the occasional 
``extreme scattering events'' and
episodes of fringes in dynamic spectra when a
line of sight passes through a particular 
discrete density enhancement.  New theoretical 
work is needed to quantify the expected statistics
of these propagation events. It is likely that numerical
modeling will be necessary to model non-stationary 
scattering media.

\noindent \bf Acknowledgements:  \rm this work was supported by the NSF under grant AST-9414144.

%
%

\appendix

\section{THEORY FOR REFRACTIVE SCINTILLATION INDEX}

\subsection*{Screen Geometry}

Here we derive expressions for the refractive
modulation index, $m_R$, as a function of the normalized diffractive
decorrelation bandwidth, $\Delta\nu_d/\nu$, for the inner-scale
and $\beta=4$ models.  The simple Kolmogorov model is included
as a special case where the inner scale goes to zero.
A general integral expression for the refractive scintillation 
index is given by equation (\ref{eq:crextmed}). With the 
medium concentrated
into a thin layer (thickness $\Delta z$ at $z_p$ from the source)
and an assumption of isotropy in the density spectrum,
the equation becomes:
\begin{eqnarray}
m_R^2 & = &\frac{16 \pi^2 r_{\rm e}^2 c^2 \Delta z}{\nu^2}
\int_0^{\infty} 
P_{N_e}(\kappa)
\exp\left[-D_{\phi}(r_{\rm F,scr}^2 \kappa)\right] \nonumber \\
& & \times \sin^2\left(\frac{r_{\rm F,scr}^2 \kappa^2}{2}\right)
\kappa \, d\kappa
\, \mbox{.}
\label{eq:app:mrscreen}
\end{eqnarray}
For both the inner-scale and $\beta=4$ 
models, the exponential function in equation (\ref{eq:app:mrscreen}) 
may be approximated by the Gaussian (Coles et al.\ 1987):  
\begin{equation}
\exp\left[ -D_{\phi}(r_{\rm F,scr}^2 \kappa) \right]
\simeq
\exp\left[ -(r_{\rm F,scr} u_{\rm scr})^2 \kappa^2 \right]
\, \mbox{,}
\label{eq:app:gauss}
\end{equation}
where  $r_{\rm F,scr}$ is the Fresnel scale and $u_{\rm scr}$ is the 
strength of scattering, as defined in section \ref{sec:mr}.
We now substitute models for the density spectrum $P_{N_e}(\kappa)$, 
given by equations (\ref{eq:extpowerlaw}) (with $\kappa_o = 0$)
and (\ref{eq:beta4}) and obtain
relations between $m_R$ and the scattering measure 
$SM = C_{N_e}^2 \Delta z$.  To these we add the 
connections between $SM$ and
$s_{\rm 0,scr}$, for each spectral model, from 
the equations for the phase structure functions (17), 
(11), and (21) of LR99.

For the inner-scale model, we obtain:
\begin{eqnarray}
\frac{8 \pi^2 r_{\rm e}^2 c^2 SM}{\nu^2} & = &
\frac{\alpha 2^{\alpha}} {s_{\rm 0,scr}^{\alpha}}
\frac{\Gamma\left(1 + \alpha/2 \right)}
     {\Gamma\left(1 - \alpha/2 \right)} \times  \nonumber \\
& & \left[
1 + \left(\frac{L_i}{s_{\rm 0,scr}}\right)^{\alpha} \mu_{\alpha}
\right]^{(2-\alpha)/\alpha}
\, \mbox{,}
\label{eq:app:sminscale}
\end{eqnarray}
where
\begin{equation}
\mu_{\alpha} = 
\left[
\frac{\alpha}{2}
\Gamma\left(1 + \frac{\alpha}{2}\right)
\right]^{\alpha/(\alpha-2)}
\, \mbox{.}
\label{eq:app:mualpha}
\end{equation}
Then substituting for $SM$ we obtain an integral for $m_R^2$,
which can be solved analytically using 
standard techniques (see e.g.\ Appendix 2 of Rickett 1973):
\begin{eqnarray}
m_R^2 & = &
2^{\alpha}
\Gamma\left( 1 + \frac{\alpha}{2}\right)
\left[
1 + \left(\frac{L_i}{s_{\rm 0,scr}} 
\right)^{\alpha} \mu_{\alpha}
\right]^{(2-\alpha)/\alpha} \times \nonumber \\
& & \left[ (\cos\psi)^{-\alpha/2}
\cos(\psi\alpha/2) - 1
\right]  \, \times \nonumber \\ 
& &
\left[ (L_i/2s_{\rm 0,scr})^2 + u_{\rm scr}^4 \right]^{\alpha /2}
\, \mbox{,}
\label{eq:app:mrscreeninscale}
\end{eqnarray}
where $u_{\rm scr} = r_{\rm F,scr}/s_{\rm 0,scr}$, 
and $\psi$ is given by:
\begin{equation}
\psi = \cot^{-1} \left[
\left( \frac{L_i}{2 s_{\rm 0,scr}u_{\rm scr}} \right)^2  
+ u_{\rm scr}^2 \right]
\, \mbox{.}
\label{eq:app:thetae}
\end{equation}
We see that $m_R$ depends on $u_{\rm scr}$ with 
$L_i/s_{\rm 0,scr}$ 
as a parameter.
Similarly the diffractive decorrelation bandwidth, 
$\Delta\nu_d/\nu$, depends on $u_{\rm scr}$ through 
equation (\ref{eq:Delta_nu_d}) with $v_d$ as a parameter, 
which is a very slow function of $L_i/s_{\rm 0,scr}$
as discussed in section \ref{sec:decbw}.

For the simple power-law model we set $L_i = 0$, 
and in the strong scattering limit ($u_{\rm scr}>>1$), 
equation (\ref{eq:app:thetae}) gives 
$\psi \approx u_{\rm scr}^{-2}$,
and we obtain the asymptotic expression:
\begin{equation}
{\displaystyle
m_R^2 = \frac{\alpha 2^{\alpha}}{4}
\left(1 - \frac{\alpha}{2}\right)
\Gamma\left(1 + \frac{\alpha}{2}\right)
\left(\frac{1}{u_{\rm scr}^2}\right)^{2-\alpha}
}
\, \mbox{.}
\label{eq:app:mrscreenkol}
\end{equation}
Putting this together for the simple Kolmogorov model 
with $\alpha=5/3$, we have the simple relation: 
$m_R = 0.459 (\Delta\nu_d/\nu)^{0.167}$, plotted as the 
thick dashed lines in the screen plots in Figures
\ref{fig:mr4} and \ref{fig:mrin}. 

For the general inner-scale model, with $L_i$ non-zero, 
we show computed curves in Figure \ref{fig:mrin}.  We can 
recognize two regimes.  Consider first the case of small $L_i$. 
With sufficiently strong scattering, 
$s_{\rm 0,scr} < L_i  < r_{\rm F,scr}$, 
$m_R^2 \propto (L_i/u_{\rm scr} r_{\rm F,scr})^{2-\alpha}$.
However, there is a complication in portraying this behavior in a plot
versus $\Delta \nu_d/\nu$ for fixed inner scale, because
of the variation of $r_{\rm F,scr}$ with distance, which is not
specified by the horizontal variable $\Delta \nu_d/\nu$.  
We deal with this by obtaining an approximate scaling of
$r_{\rm F,scr}$ with $\Delta \nu_d/\nu$.  When 
$s_{\rm 0,scr} < L_i$, equation (\ref{eq:app:sminscale}) 
relates $SM \propto s_{\rm 0,scr}^{-2}$, or
$SM r_{\rm F,scr}^2 \propto u_{\rm scr}^{2}$.  
We now argue that on average 
$SM \propto$ pulsar distance $\propto r_{\rm F,scr}^2$,
and eliminating $SM$, we see that, 
$r_{\rm F,scr} \propto u_{\rm scr}^{0.5}$.  With these 
scalings and $s_{\rm 0,scr} < L_i  < r_{\rm F,scr}$, 
we see:
\begin{eqnarray}
m_R^2 & \propto & (L_i/u_{\rm scr} r_{\rm F,scr})^{2-\alpha}
\propto (L_i/r_{\rm F,ref} u_{\rm scr}^{1.5} )^{2-\alpha} \nonumber \\
& \propto & u_{\rm scr}^{-0.5} \propto  (\Delta \nu_d/\nu)^{0.25}
\, \mbox{.}
\label{eq:mr.asymptote1}
\end{eqnarray}
Here $r_{\rm F,ref}$ is the Fresnel scale for a ``reference''
pulsar, and the last version in equation (\ref{eq:mr.asymptote1})
comes from using the Kolmogorov exponent $\alpha = 5/3$, which 
gives the asymptotic slope of 1/8 for curves at the lower left of 
Figure \ref{fig:mrin}.  Now with $ L_i$ and $r_{\rm F,scr}$ fixed, 
let $s_{\rm 0,scr}$ increase (i.e.\ less scattering) until 
$L_i  < s_{\rm 0,scr} < r_{\rm F,scr}$.  The inner scale is no 
longer important, and we get the same relation
as for the simple Kolmogorov spectrum: 
$m_R^2 \propto u_{\rm scr}^{-2(2-\alpha)} 
\propto (\Delta \nu_d/\nu)^{1/3}$; visible where the curves
steepen with increasing $\Delta \nu_d/\nu$.

For larger inner scales, the curves in Figure \ref{fig:mrin}
show a pronounced peak. These occur for inner scales
greater than the Fresnel scale and will be associated with
focusing and caustics.  Since our treatment only includes
the first order term in the low wavenumber expansion,
it is not reliable in the region of the peak.
Goodman et al.\ (1987) have discussed caustics
at length for the same inner-scale spectrum.  They note that
when $s_{\rm 0,scr}  < r_{\rm F,scr} < L_i  < s_R$,
the scintillation power spectrum starts to fill in at scales
intermediate between the diffractive and refractive 
wavenumbers.  Their equation (2.5.7) gives an estimate of 
the variance in this extra term as:
\begin{eqnarray}
m_{\rm int}^2 & \sim &  2 (L_i/s_R)^2 \ln (L_i/s_{\rm 0,scr}) \nonumber \\
& \sim & 2  (L_i/u_{\rm scr} r_{\rm F,scr})^2 
\ln (L_i u_{\rm scr}/r_{\rm F,scr})
\mbox{.}
\label{eq: mr.asymptote2}
\end{eqnarray}
With fixed $L_i$,  $m_{\rm int}^2$ increases much more
steeply with decreasing $u_{\rm scr}$ than does $m_R^2$.
Thus as $u_{\rm scr} r_{\rm F,scr}$ decreases, 
there are fewer and fewer independent  phase perturbations 
across the scattering disc.  When $u_{\rm scr} r_{\rm F,scr}
\sim L_i$, focusing represented by $m_{\rm int}^2 
\sim 4 \ln(L_i/r_{\rm F,scr}) > 1$ occurs.
We note that the theoretical $m_R$ versus $\nu_d/\nu$
plots in Gupta et al.\ (1993) omit the
focusing condition and are wrong for inner scales
greater than the Fresnel scale.  The effect of higher order
terms in the low-wavenumber expansion has been studied
by Dashen \& Wang (1993).  They obtain a more efficient
expansion scheme that gives improved accuracy near the peak
in scintillation index.  Nevertheless, it seems that a
reliable prediction for the behavior near the
peak in scintillations requires numerical evaluation.
This becomes even more necessary in treating an extended 
scattering medium.

We also note that the drop in 
$m_R$ as scattering gets weaker past the peak
in Figure \ref{fig:mrin} is real. It represents
the fact that when a square-law structure function
applies for scales from $s_{\rm 0,scr}$
up to the scattering disc size, there is insufficient
phase curvature and the scintillations remain weak
even though $u_{\rm scr} > 1$.  In such circumstances
the ``scattering disc'' is a misnomer,
since an observer would see only a single angle of arrival, that
could wander over a region of scale $u_{\rm scr} r_{\rm F,scr}$.

Turning to the screen analysis of the
$\beta=4$ model, we relate $s_{\rm 0,scr}$
to $SM$ using equation (21) from LR99,  where the phase structure 
function equals one. This gives the following analog of equation
(\ref{eq:app:sminscale}) for the inner-scale model:
\begin{equation}
\frac{\pi^2 r_{\rm e}^2 c^2 SM}{\nu^2} =
\left\{
s_{\rm 0,scr}^2 \ln\left[1 + 4 \left(
\frac{L_o}{s_{\rm 0,scr}}\right)^2 \right]
\right\}^{-1}
\mbox{.}
\label{eq:app:smbeta4}
\end{equation}
Substituting the $\beta=4$ model for the density spectrum
in equation (\ref{eq:app:mrscreen}) and letting
$\eta = \kappa^2$, we obtain:
\begin{eqnarray}
m_R^2 \; = \; \frac{8 \pi^2 r_{\rm e}^2 c^2 SM}{\nu^2} \;
\int_0^{\infty} 
\frac{1}{(\eta + \kappa_o^2)^2} \; \times \nonumber \\
\exp [ -(r_{\rm F,scr} u_{\rm scr})^2\eta ] \;
\sin^2 ( r_{\rm F,scr}^2 \eta/2 ) \; d\eta
\label{eq:app:mrscreenb4_1}
\end{eqnarray}
In order to evaluate this integral, we
let $g(\eta)$ represent the integrand, and we let $g_1(\eta)$ 
be the same as $g(\eta)$ but with $\kappa_o$ set equal to zero.
We can then rewrite the integral in equation 
(\ref{eq:app:mrscreenb4_1}) as:
\begin{eqnarray}
& &  m_R^2 \;\; = 8 
\left\{
s_{\rm 0,scr}^2 \ln\ [1 + (2 L_o/s_{\rm 0,scr})^2 ]
\right\}^{-1}  \times \nonumber \\
& & \left\{
\int_0^{\infty} g_1(\eta)  d\eta -
\int_0^{\infty} [g_1(\eta) - g(\eta) ] d\eta
\right\}
\label{eq:app:mrscreenb4_2}
\end{eqnarray}
The first integral can be evaluated analytically (see e.g.\ 
\mbox{Gradshteyn} \& \mbox{Ryzhnik} 1965).
In strong scattering, the exponential term cuts off
the oscillations of the sine term, which can be approximated
by its argument, and the $g_1(\eta) - g(\eta)$
becomes negligible for values of $\eta$ larger than $\kappa_o^2$.
With these approximations we can also do the second integral. 
Putting these together, we obtain:
\begin{eqnarray}
& & m_R^2  = 
\left\{ \ln [ 1 + (2 L_o/s_{\rm 0,scr})^2 ]  \right\}^{-1} \times \nonumber \\
& & \left\{ 4 u_{\rm scr}^2 \tan^{-1}(u_{\rm scr}^{-2}) -
2 u_{\rm scr}^4\ln (1 + u_{\rm scr}^{-4} ) -  \right. \nonumber \\ 
& & \left. 2 \zeta \, [(2 + \zeta) \, e^{\zeta} \, {\rm E_1}(\zeta) - 1 ]
\right\}
\, \mbox{,}
\label{eq:app:mrscreenb4}
\end{eqnarray}
where ${\rm E_1}$ is the exponential integral,
$\zeta = (s_0\kappa_o)^2 u_{\rm scr}^4 = (s_R \kappa_o)^2$.  
In the latter form $s_R = r_{\rm F,scr} u_{\rm scr}$ is the 
refractive scale (equal to scattering disk radius).
Again, $m_R$ is related to $\Delta\nu_d/\nu$ through: 
$\Delta\nu_d/\nu = v_d/u_{\rm scr}^2$. The resulting curves are 
shown in the screen panels of Figure \ref{fig:mr4}. 
Consider the asymptotic behavior for $m_R^2$
in equation (\ref{eq:app:mrscreenb4}) as a function of
$u_{\rm scr}$. As the strength of scattering
$u_{\rm scr}$ increases,  $\Delta\nu_d/\nu$ 
decreases ($\propto u_{\rm scr}^{-2}$).  In equation 
(\ref{eq:app:mrscreenb4_1}), the exponential term
cuts off the integral (at $1/s_R^2$) before the oscillations of the
sin$^2$ Fresnel filter, which then approximates 
$\eta^2 r_{\rm F,scr}^4/4$.  If also
$\zeta = (s_R \kappa_o)^2 \ll 1$, we can ignore $\kappa_o^2$
in the denominator and the remaining $\eta^{-2}$ cancels the 
$\eta^2$ from the Fresnel filter, and the integral depends 
only on $1/s_R$. In this approximation, $m_R^2$ is then simply
proportional to the slowly varying logarithmic term,
which explains the relatively flat part of the curves
in Figure \ref{fig:mr4}; under these conditions,
in equation (\ref{eq:app:mrscreenb4})
the first two terms in the curly brackets sum to 2
and the last term is negligible.  With $L_o$ fixed, now let
$u_{\rm scr}$ increase, making $s_R$ increase. Eventually
$\zeta$ becomes greater than one when the scattering disk
becomes greater than the outer scale. At this point the
exponential term cuts off the integral below 
$\kappa_o^2$, where the spectrum flattens.
As $u_{\rm scr}$ increases even further,
the integral decreases steeply, causing the down-turn 
at very small $\Delta\nu_d/\nu$.   We again note that
our expressions rely on the first order of an expansion
and will not be reliable near the peak in the scintillation
index.  However, there is not the same focusing
condition that applied for very large inner scales.

\subsection*{Extended Scattering Medium}

In order to obtain expressions for $m_R$ for the two spectrum 
models in the \it uniform \rm extended medium geometry, we 
must complete the line-of-sight integrals in equation
(\ref{eq:crextmed}) in addition to following the steps
used in the screen geometry. 
For each distance $x = z/L$ in the line of sight, there is also
an integration over variable $y$ in the exponential cut-off.  
If $D'(s) \propto s^{\alpha}$, this $y$-integration yields  
$L D'(\kappa r_{F,L}^2 x(1-x))/(\alpha+1)$, where 
$r_{{\rm F},L}= \sqrt{Lc/(2\pi\nu)}$.  This again provides
a low-pass cut-off at the reciprocal of the radius of the effective 
scattering disk, where $\kappa \sim (u_{S}r_{{\rm F},L})^{-1}$.
We define the scattering strength by 
$u_S = r_{{\rm F},L}/s_{0_S}$, with the
field coherence scale $s_{0_S}$, as in LR99, defined where 
the spherical wave structure function equals unity,
measured in the observing plane.
For the other spectrum  models there is not
such a simple relation for the $y$-integration, 
but there is still an effective cutoff given by a similar
equation.  When the $x$-integration is completed, the 
effective Fresnel scale is actually smaller than
$r_{{\rm F},L}$ due to averaging over the $\sqrt{x (1-x)}$.

In analogy with the screen geometry, we make use of identities
similar to those given by equations (\ref{eq:app:sminscale}) and 
(\ref{eq:app:smbeta4}).  For the extended medium, these identities
are in turn derived from the {\em wave} structure functions for
the inner-scale and $\beta=4$ models (cf.\ LR99).  The identities
obtained thus will be similar to those for the screen geometry,
except that for the inner-scale model, (1) there will be a factor 
of 3 on the right side of equation (\ref{eq:app:sminscale}), and
(2) the 1 in square brackets is replaced by 
$[3 / (1+\alpha)]^{\alpha/(\alpha-2)} \approx 1.8$
for the Kolmogorov exponent.  For the
$\beta=4$ model, the only change will be a factor of 3 on the
right side of equation (\ref{eq:app:smbeta4}).  
We use all of the aforementioned identities
for the inner-scale and $\beta=4$ models and compute
the $x$ integral numerically, since it cannot be carried out
analytically.  

The shapes of the curves bear a close relationship to the screen
results, though the extended medium values
generally lie above the associated screen values at
the same $\Delta \nu_d/\nu$.

\section{Diffractive Intensity Correlation Function}

The second moment of intensity is needed to describe
the fluctuations of intensity.  Under 
strong scintillation conditions, separate forms can be 
used for refractive and diffractive fluctuations, since
their spatial scales differ by several orders 
of magnitude. In LR99, as elsewhere in the ISS literature, the 
correlation of diffractive scintillations is approximated by
the squared magnitude of the second moment of the field,
leading to the simple result that the spatial scale of the diffractive
scintillations is equal to the scale where the phase structure function
equals unity ($s_0$). However, Goodman and Narayan (1985) showed that 
for steep spectra ($\beta>4$) this is no longer the case and
the diffractive scale can be larger than $s_0$. 
Here we examine this question for the $\beta=4$ 
spectrum.  We give the details for a phase screen
with plane wave source, which are readily generalized to
a spherical wave source.

The two-frequency intensity cross-spectrum at wave\-number 
{\boldmath$ \kappa$} for a screen at distance $L$ is given by the 
Fourier-like integral equation (17) of CCFFH. 
This  depends on the combination of structure functions $V_4$,
which for a plasma screen can be written as:
\begin{eqnarray}
V_4 = \frac {k_m^2}{k_1^2} D_{\phi}(\mbox{\boldmath $\kappa$} \frac{L}{k_1})
+ \frac {k_m^2}{k_2^2} D_{\phi}(\mbox{\boldmath $\kappa$} \frac{L}{k_2})
-  D_{\phi}( \frac{\mbox{\boldmath$\kappa$}L}{k_o} + \mbox{\boldmath$\beta $}')\nonumber \\
-  D_{\phi}( \frac{\mbox{\boldmath$\kappa$}L}{k_o} - \mbox{\boldmath$\beta $}') 
+ D_{\phi}(\mbox{\boldmath$\beta $}' +  \frac{\mbox{\boldmath$\kappa$} \epsilon L}{k_o})
+ D_{\phi}(\mbox{\boldmath$\beta $}' -  \frac{\mbox{\boldmath$\kappa$} \epsilon L}{k_o}) \, ,
\label{eq:v4}
\end{eqnarray}
where $k_1$ and $k_2$ are the two radio wavenumbers,
$k_m$ is their geometric mean, $\bar{k}$ is their arithmetic mean,
$k_o = k_m^2/\bar{k}$, $\epsilon=|k_1-k_2|/2\bar{k}$,
and $D_{\phi}$ is evaluated at $k_m$; $\mbox{\boldmath$\beta $}'$
is a spatial offset which is the variable of integration.  

Consider first the single frequency case $k_1=k_2 (\epsilon=0$).
In the limit of very large $\kappa$, the first four structure functions
saturate and sum to zero.  The last two are equal and
$V_4 \approx 2 D_{\phi}(\beta ')$,
which gives the simple diffractive limit mentioned above.  
This is the zero-order term of an expansion, which is obtained
in terms of the sum of the first four terms as a small quantity.
The zero order result requires full saturation, 
which requires $\kappa L/k_m \simgreat L_o$.  In diffractive
scintillation $\kappa \sim 1/s_0$; hence, the condition 
becomes that the refractive scale $s_R = L/(k_m s_0) \simgreat L_o$. 
Our concern here is to consider what happens
when the diffractive $\kappa$ is not large enough for 
saturation of $D_{\phi}$.  For shallow density spectra
($\beta < 4$), small argument approximations to the structure function
follow an exponent $\beta-2 < 2$, and the zero-order term gives
a good approximation even when $s_R < L_o$. However, for
steep spectra Goodman and Narayan (1985) showed that the
leading term in the structure function follows a square law,
which exactly cancels in $V_4$; the result is that
the high wavenumber limit depends on the next term
in the structure function expansion, which yields a diffractive scale 
that is greater than the scale $s_0$ (defined by the square
law term).

Now we consider the case for $\beta=4$ model, when the scattering disc
$s_R$ is smaller than the outer scale $L_o$. Here 
we can approximate equation (\ref{eq:struc4}) by: 
\begin{equation}
D_{\phi}(s) = \frac {s^2}{s_0^2} - \frac {s^2 \ln(s^2/s_0^2)}
{s_0^2 \ln(4/s_0^2\kappa_o^2)}
\, \mbox{.}
\label{eq:Dapprox}
\end{equation}
As for the steep spectra, the leading term 
in the structure function follows a square law,
which cancels when substituted into equation (\ref{eq:v4}).  
$V_4$ can then be approximated for large $\kappa$
by expanding in $\beta ' k_m/(\kappa L)$.  The result is:
\begin{equation}
V_4 \sim \frac {2 \beta '^2[\ln(u^4) + 1 + 2 \cos^2\theta - \ln(\beta '^2/s_0^2)]}
{s_0^2 \ln(4/s_0^2\kappa_o^2)}
\, \mbox{.}
\label{eq:v4approx}
\end{equation}
Here $u$ is the strength of scattering defined 
in equation (\ref{eq:Delta_nu_d}); $\theta$ is the angle between vectors
{\boldmath $\kappa$} and $\mbox{\boldmath $\beta$}'$. 
We note that for a Kolmogorov spectrum in the high-wavenumber limit,
$V_4$ also includes terms in $\cos^2\theta$, which would
be accounted for in the higher order terms of the expansion.
A result similar to equation (\ref{eq:v4approx})
is given by Dashen \& Wang (1993), though
for a 1-dimensional phase screen.
In considering the spectrum of intensity fluctuations 
(eq.\ 17 of CCFFH), one can show that the dominant wavenumber 
is approximately $\kappa \sim \beta '^{-1}$
With these substitutions in $V_4$, which we then set $=2$, we 
solve for $\beta '$; this gives an 
approximate equation for the diffractive spatial scale $s_d$: 
\begin{equation}
s_d^2 \sim s_0^2 \frac {\ln(4/s_0^2\kappa_o^2)}{\ln(u^4)+ 2}
\, \mbox{.}
\end{equation}
Under the condition assumed in this approximation,
$s_R \ll L_o$, we find $s_d > s_0$.  However,
in practice the ratio $s_d/s_0$
never becomes large.  With a large outer scale,
say $L_o = 3$~pc, and typical observing conditions
$s_0 \sim 10^8$ m and $u \simgreat 100$, we find $s_d \simless 1.7 s_0$.
Thus the diffractive scale could be 70\% greater than
$s_0$ and would slowly approach $s_0$ for smaller outer scales.

Turning to the two-frequency intensity correlation
($0 < \epsilon \ll 1$) in the diffractive limit of
large $\kappa$, the results of CCFFH still apply.
Namely, that the last two terms of
equation (\ref{eq:v4}) largely control the decorrelation
versus frequency. They group the remaining terms
into a filter that depends only on
$\kappa$ and a smaller term that becomes the basis of the expansion. 
The filter term was discussed by LR99 and shown to be important
only as the strength of scattering decreases.
It is the last two terms in $V_4$ that determine the zero-order
result, so we looked at the effect of the higher order terms.
The quantity that we are ultimately concerned with
is the cross-correlation of intensity at offset
frequencies at the same observing point.  This comes
from the integral of the cross-spectrum.  Equations (31) through (34)
of CCFFH give the zero and first order terms 
of the cross spectrum in terms of the spectrum of 
refractive index fluctuations in the layer.  
For the $\beta=4$ spectrum we reduced these to a 
sum of confluent hypergeometric functions, which can
be explicitly computed. For a sample observing condition
we found that
the higher order terms for the
cross-spectrum itself were significant compared 
to the zero-order term; however when integrated to give
intensity cross-correlation, they only had a  
minor effect on the decorrelation bandwidth itself
($\simless 5$\% increase).  The reason for this appears to be the
dominant effect of the last two terms in the $V_4$ summation
with unequal frequencies.

To summarize we find a modest (logarithmic) increase in the diffractive
scale relative to the field coherence scale $s_0$, but that
this remains less than a factor of 1.7 for the likely ISS
parameters. This is accompanied by a smaller increase
in the decorrelation bandwidth relative to the calculations
of LR99, which relied on the normal zero-order expansion
at high wavenumbers. This small offset in the decorrelation
bandwidth is negligible compared to the measurement errors
for the observations under consideration.  We assume that the
conclusions reached here for a screen would also apply for 
an extended scattering medium.

%


\begin{thebibliography}{}


\bibitem[Armstrong et al.\ 1995]{Armstrong95}
  Armstrong, J. W., Rickett, B. J., \& Spangler, S. R. 1995,
  \apj, 443, 209

\bibitem[Bhat et al. 1998]{Bhat98}
  Bhat, N. D. R., Gupta, Y., \& Rao, A. P. 1998, 
  \apj, 500, 262

\bibitem[BGR 1999a]{BRG99a}
  Bhat, N. D. R., Rao, A. P., \&  Gupta, Y. 1999a, 
  \apjs, 121, 483

\bibitem[BGR 1999a]{BRG99a}
  Bhat, N. D. R., Rao, A. P., \&  Gupta, Y. 1999b, 
  \apj, 514, 249

\bibitem[Bhat et al. 1999b]{BGR99c}
  Bhat, N. D. R., Gupta, Y., \& Rao, A. P. 1999c, 
  \apj, 514, 272

\bibitem[Biskamp]{Biskamp93}
 Biskamp, D. 1993, \it Nonlinear Magnetohydrodynamics \rm , 
 Cambridge Univ. Press, Cambridge, UK

\bibitem[Blandford \& Narayan 1985]{Blandford85}
  Blandford, R., \& Narayan, R. 1985,
  \mnras, 213, 591

\bibitem[Codona et al.\ 1986]{Codona86}
  Codona, J. L., Creamer, D. B., Flatt\'{e}, S. M., Frehlich, R. G.,
  \& Henyey, F. S. 1986, Radio Science, 21, No.\ 5, 805 [CCFFH]

\bibitem[Cole et al.\ 1970]{Cole70}
  Cole, T. W., Hesse, H. K., \& Page, C. G. 1970,
  Nature, 225, 712

\bibitem[Coles et al.\ 1987]{Coles87}
  Coles, W. A., Frehlich, R. G., Rickett, B. J., \&
  Codona, J. L. 1987,
  \apj, 315, 666

\bibitem[Cordes et al.\ 1985]{Cordes85}
  Cordes, J. M., Weisberg, J. M., \& Boriakoff, V. 1985,
  \apj, 288, 221 [CWB]

\bibitem[Cordes, Pidwerbetsky, \& Lovelace 1986]{Cordesetal86}
  Cordes, J. M., Pidwerbetsky, A., \& Lovelace, R. V. E. 1986,
  \apj, 310, 737

\bibitem[Cordes \& Wolszczan 1986]{Cordes86}
  Cordes, J. M., \& Wolszczan, A. 1986, \apj, 307, L27

\bibitem[Cordes et al.\ 1991]{Cordes91}
  Cordes, J. M., Weisberg, J. M., Frail, D. A., Spangler, S. R., \& 
  Ryan, M. 1985, \apj, 288, 221

\bibitem[Dashen and Wang (1993)]{Dashen93}
  Dashen, R., \& Wang, G. Y. 1993, 
  J. Opt. Soc. Am. A, 10, 1219

\bibitem[Fiedler et al.\ 1987]{Fiedler87}
  Fiedler, R. L., Dennison, B., Johnston, K. J., \& Hewish, A. 1987,
  Nature, 326, 675

\bibitem[Goldreich \& Sridhar 1995]{Goldreich95}
  Goldreich, P., \& Sridhar, S. 1995, 
  \apj, 438, 763

\bibitem[Goldreich \& Sridhar 1997]{Goldreich97}
  Goldreich, P., \& Sridhar, S. 1997,
  \apj, 485, 680

\bibitem[Goodman \& Narayan 1985]{Goodman85}
  Goodman, J., \& Narayan, R. 1985, 
  \mnras, 214, 519

\bibitem[Goodman et al. 1987]{Goodman87}
  Goodman, J., Romani, R. W., Blandford, R. D., \& Narayan, R. 1987,
  \mnras, 229, 73

\bibitem[Gradshteyn \& Ryzhnik 1965]{Gradshteyn65}
  Gradshteyn, I. S., \& Ryzhnik, I. M. 1965,
  Table of Integrals, Series, and Products
  (New York: Academic Press)

\bibitem[Gupta et al.\ 1993]{Gupta93}
  Gupta, Y., Rickett, B. J., \& Coles, W. A. 1993,
  \apj, 403, 183

\bibitem[Gupta et al.\ 1994]{Gupta94}
  Gupta, Y., Rickett, B. J., \& Lyne, A. G. 1994,
  \mnras, 269, 1035

\bibitem[Gupta et al.\ 1999]{Gupta99}
 Gupta, Y., Bhat, N. D. R., \& Rao, A. P. 1999
  \apj, in press 

\bibitem[Helfand et al.\ 1977]{Helfand77}
  Helfand, D. J., Fowler, L. A., \& Kuhlman, J. V. 1977,
  \aj, 82, 701

\bibitem[Higdon 1984]{Higdon84}
  Higdon, J. C. 1984,
  \apj, 285, 109

\bibitem[Higdon 1986]{Higdon86}
  Higdon, J. C. 1986,
  \apj, 309, 342

\bibitem[Johnston, Nicastro, \& Koribalski 1998]{Johnston98}
  Johnston, S., Nicastro, L., \& Koribalski, B. 1998,
  \mnras, 297, 108

\bibitem[Kaspi \& Stinebring 1992]{Kaspi92}
  Kaspi, V. M., \& Stinebring, D. R. 1992,
  \apj, 392, 530

\bibitem[Lambert 1998]{Lambert98}
  Lambert, H. C. 1998, Interstellar Electron Density Spectra,
  PhD Thesis, University of California, San Diego

\bibitem[Lambert \& Rickett 1999]{Lambert1999}
  Lambert, H. C., \& Rickett, B. J. 1999, 
  submitted to \apj [LR99]

\bibitem[Lestrade et al 1998]{Lestrade1998}
 Lestrade, J-F, Rickett, B. J., \& Cognard, I. 1998, 
 A\&A, 334, 1068 

\bibitem[Minter \& Spangler 1997]{Minter97}
  Minter, A. H., \& Spangler, S. R. 1997,
  \apj, 485, 182

\bibitem[Narayan 1992]{Narayan92}
    Narayan, R. 1992, Phil. Trans. R. Soc. London A, 341, 151

\bibitem[Papoulis 1991]{Papoulis91}
  Papoulis, A. 1991, Probability, Random Variables, and Stochastic
  Processes, Third Edition (New York: McGraw-Hill Inc.)

\bibitem[Pouquet 1978]{Pouquet78}
  Pouquet, A. 1978, J. Fluid Mech., 88, 1

\bibitem[Prokhorov et al.\ 1975]{Prokhorov75}
  Prokhorov, A. M., Bunkin, F. V., Gochelashvily, K. S.,
  \& Shishov, V. I. 1975, Proc.\ IEEE, 63, 790

\bibitem[Ratcliffe 1956]{Ratcliffe56}
  Ratcliffe, J. A. 1956,
  Reports on Progress in Physics 19, 188

\bibitem[Rickett 1973]{Rickett73}
  Rickett, B. J. 1973, \jgr, 78, 1543

\bibitem[Rickett 1977]{Rickett77}
  Rickett, B. J. 1977,
  \araa, 15, 479

\bibitem[Rickett et al. 1984]{Rickett84}
  Rickett, B. J., Coles, W.A., \& Bourgois, G. 1984,
  \aap, 134, 390

\bibitem[Rickett \& Lyne 1990]{RickettLyne90}
  Rickett, B. J., \& Lyne, A. G. 1990, \mnras, 244, 68

\bibitem[Rickett et al.\ 1997]{Rickett97}
  Rickett, B. J., Lyne, A. G., \& Gupta, Y. 1997,
  \mnras, 287, 739

\bibitem[Rickett et al.\ 1999]{Rickett99}
  Rickett, B. J., Coles, W. A., \&  Markkanen, J. 1999,
  submitted to \apj

\bibitem[Roberts \& Ables 1982]{Roberts82}
  Roberts, J. A., \& Ables, J. G. 1982,
  \mnras, 201, 1119

\bibitem[Romani et al.\ 1986]{Romani86}
  Romani, R. W., Narayan, R., \& Blandford, R. 1986,
  \mnras, 220 19

\bibitem[Smirnova et al 1998]{Smirnova98}
 Smirnova, T. V., Shishov, V. I., \& Stinebring, D. R. 1998,
 Astronomical Reports, 42, 766

\bibitem[Spangler \& Gwinn 1990]{Spangler90}
  Spangler, S. R., \& Gwinn, C. R. 1990, \apj, 353, L29

\bibitem[Spangler 1991]{Spangler91}
  Spangler, S. R. 1991, \apj, 376, 540

\bibitem[Spangler 1999]{Spangler99}
  Spangler, S. R. 1999, \apj, 522, 879

\bibitem[Sridhar \& Goldreich 1994]{Sridhar84}
  Sridhar, S., \& Goldreich, P. 1994,
  \apj, 432, 612

\bibitem[Stinebring, 1996]{Stinebring96}
 Stinebring, D. R., Smirnova, T. V., Hovis, J., Kempner, J. C.,
 Myers, E. B., Hankins, T. H., Kaspi, V. M., \& Nice, D. J. 1996,
 in ``Pulsars: Problems and Progress'', Ed.\ Johnston, S., 
 Walker, M. A., \& Bailes, M., Proceedings of IAU Colloquium 160,
 ASP Conference Series, 105, 455

\bibitem[Tatarskii 1961]{Tatarskii61}
  Tatarskii, V. I. 1961, Wave Propagation in a Turbulent Medium
  (New York: Dover Publications)

\bibitem[Taylor \& Cordes 1993]{TC93}
 Taylor, J. H., \& Cordes, J. M. 1993,
 \apj, 411, 674

\bibitem[Taylor et al 1993]{Tayloretal93}
 Taylor, J. H., Manchester, R. N., \& Lyne, A. G. 1993, 
 \apjs, 88, 529

\bibitem[Trotter et al 1998]{Trotter98}
 Trotter, A. S., Moran, J. M., \&  Rodr\'{i}guez, L. F. 1998,
 \apj, 493, 666


\end{thebibliography}
\end{document}